\documentclass[aps,prd,onecolumn,showpacs,showkeys,superscriptaddress,preprintnumbers,floatfix,reprint]{revtex4}
\usepackage{graphicx}
\usepackage{amsmath}
\usepackage{bm}
\usepackage{yhmath}
\usepackage{hyperref}
\usepackage{subfigure}
\usepackage{color}
\usepackage{cases}
\usepackage{subfigure}
\usepackage{dcolumn,booktabs,bm}

\usepackage{amsfonts,amssymb,stmaryrd,latexsym,amsmath}
\usepackage{textcomp}

\usepackage{array} % for extrarowheigh
\newcounter{RomanNumber}

\newcommand{\lyxmathsym}[1]{\ifmmode\begingroup\def\b@ld{bold}
  \text{\ifx\math@version\b@ld\bfseries\fi#1}\endgroup\else#1\fi}

\def\rmI{{\rm I}}
\def\rmII{{\rm I\!I}}
\def\rmIII{{\rm I\!I\!I}}

\def\mud{{\lambda}}

\allowdisplaybreaks
\UseRawInputEncoding

\begin{document}
%\preprint{}
\title{The magnetic moments and electromagnetic form factors of the decuplet baryons in chiral perturbation theory}

\author{Hao-Song Li}\email{haosongli@pku.edu.cn}\affiliation{School of Physics and State Key Laboratory of Nuclear Physics and Technology, Peking University, Beijing 100871, China}

\author{Zhan-Wei Liu}\email{zhan-wei.liu@adelaide.edu.au}
\affiliation{School of Physical Science and Technology, Lanzhou University, Lanzhou 730000, China}
\affiliation{CSSM, Department of Physics, University of Adelaide, Adelaide SA 5005, Australia}

\author{Xiao-Lin Chen}\email{chenxl@pku.edu.cn}
\affiliation{School of Physics, Peking University, Beijing 100871,
China}

\author{Wei-Zhen Deng}\email{dwz@pku.edu.cn}
\affiliation{School of Physics, Peking University, Beijing 100871,
China}

\author{Shi-Lin Zhu}\email{zhusl@pku.edu.cn}\affiliation{School of Physics and State Key Laboratory of Nuclear Physics and Technology, Peking University, Beijing 100871, China}\affiliation{Collaborative Innovation Center of Quantum Matter, Beijing 100871, China}

\begin{abstract}

We have systematically investigated the magnetic moments and
magnetic form factors of the decuplet baryons to the
next-to-next-leading order in the framework of the heavy baryon
chiral perturbation theory. Our calculation includes the
contributions from both the intermediate decuplet and octet baryon
states in the loops. We have also calculated the charge and magnetic
dipole form factors of the decuplet baryons. Our results may be
useful to the chiral extrapolation of the lattice simulations of the
decuplet electromagnetic properties.

\end{abstract}

% 13.25.Gv  Decays of J, and other quarkonia
% 14.40.Pq  Heavy quarkonia
% 13.75.Lb meson-meson interactions

%{\it PACS}: ~13.25.Gv, ~13.75.Lb\\


\maketitle

\thispagestyle{empty}

%\vspace{3mm}

%%%%%%%%%%%%%%%%%%%%%%%%%%%%%%%%%%%%
\section{Introduction}\label{Sec1}
%%%%%%%%%%%%%%%%%%%%%%%%%%%%%%%%%%%%

Chiral perturbation theory (ChPT) is a very useful framework in
hadron physics in the low energy regime. ChPT was first proposed to
study the purely pseudoscalar meson system with the consistent
chiral power counting scheme ~\cite{Weinberg:1978kz}, which enables
us to calculate either a physical process or hadron property order
by order. For example, the pion pion scattering amplitude in the low
energy regime can be expanded in terms of ${m_\pi\over
\Lambda_\chi}$ and ${p\over \Lambda_\chi}$ where $\Lambda_\chi=4\pi
f_\pi$ and $p$ is the three-momentum of the pion. In the chiral
limit, $m_\pi \to 0$. The above scattering amplitude converges
quickly with the soft pion momentum.

The extension of the ChPT to the matter field introduces a new large
energy scale, the mass of the matter field which does not vanish in
the chiral limit. Hence this mass scale $M$ will spoil the
convergence of the chiral expansion. To overcome this obstacle, the
heavy baryon chiral perturbation theory (HBChPT) was developed
~\cite{Jenkins:1990jv,Bernard:1992qa}. Within this scheme, one also performs the
heavy baryon expansion in terms of $1/M$ together with the chiral
expansion. With the help of HBChPT, the octet baryon masses, Compton
scattering amplitudes, axial charge, various electromagnetic form
factors and many other observables have been investigated
systematically ~\cite{Bernard:1992qa,Bernard:1995dp,Bernard:1993nj,Holstein:1992xr,
Bernard:1996gq,Mojzis:1997tu,Fettes:1998ud,Fettes:2001cr,Shanahan:2014cga}.

However, because of the non-relativistic treatment of the baryon
propagators, HBChPT also has its shortcomings. To satisfy the
analyticity constraints lost in the HBChPT, the covariant ChPT has
been applied to the study of several physical observables such as
the pion scattering, baryon magnetic moments and axial form factors,
baryon
masses~\cite{Gegelia:1999gf,Fuchs:2003qc,Fuchs:2003ir,Lehnhart:2004vi,MartinCamalich:2010fp,Alarcon:2011zs,Ledwig:2014rfa}.
In Ref~\cite{Gegelia:1999gf}, Gegelia addressed the problem of
matching HBChPT to the relativistic theory. A new renormalization
scheme leading to a simple and consistent power counting in the
single-nucleon sector of relativistic chiral perturbation theory was
discussed in Ref.~\cite{Fuchs:2003qc}. The electromagnetic form
factors of the nucleon were calculated to order $\mathcal{O}(p^{4})$
in the relativistic chiral perturbation theory in
Ref.~\cite{Fuchs:2003ir}. In Ref.~\cite{Lehnhart:2004vi}, the masses
of the ground state baryon octet and the nucleon sigma terms were
discussed in the framework of manifestly Lorentz-invariant baryon
chiral perturbation theory. An analysis of the baryon octet and
decuplet masses using covariant $SU(3)$-flavor chiral perturbation
theory up to next-to-leading order was presented in
Ref.~\cite{MartinCamalich:2010fp}. A novel analysis of the $\pi N$
scattering amplitude in Lorentz covariant baryon chiral perturbation
theory renormalized in the extended-on-mass-shell scheme have been
presented in Ref.~\cite{Alarcon:2011zs}. In
Ref.~\cite{Ledwig:2014rfa}, the octet-baryon axial-vector charges
were studied up to $\mathcal{O}(p^{3})$ using the covariant baryon
chiral perturbation theory with explicit decuplet contributions.

Covariant ChPT also has problems in the power counting introduced by
the baryon mass as a new large scale. To combine the advantages of
the relativistic and the heavy-baryon approaches, the infrared
regularization was proposed in
Refs.~\cite{Ellis:1997kc,Becher:1999he}. Kubis employed the infrared
regularization scheme to analyze the electromagnetic form factors of
the nucleon to fourth order in relativistic baryon chiral
perturbation theory in Refs.~\cite{Kubis:2000zd,Kubis:2000aa}. In
Ref.~\cite{Bernard:2003xf}, a systematic infrared regularization for
chiral effective field theories including spin$-3/2$ fields was
discussed. In Ref.~\cite{Bruns:2004tj}, the authors extended the
method of the infrared regularization to spin$-1$ fields. In
Refs.~\cite{Schindler:2003xv,Schindler:2005ke}, the authors
reformulated the infrared regularization of Becher and
Leutwyler~\cite{Becher:1999he} in a form analogous to their extended
on-mass-shell renormalization scheme and calculated the
electromagnetic form factors of the nucleon up to fourth order. In
Ref.~\cite{Alarcon:2011kh}, the authors analyzed the pion-nucleon
scattering using the manifestly relativistic covariant framework of
Infrared Regularization up to $\mathcal{O}(p^{3})$ in the chiral
expansion.

In the last two decades, there has been lots of investigations of
the baryon properties in chiral perturbation
theory~\cite{WalkerLoud:2004hf,Tiburzi:2004rh,Wang:2008vb,Camalich:2009uf,
Syritsyn:2009mx,Ahuatzin:2010ef,Ledwig:2011cx,Lensky:2009uv,Long:2009wq,Birse:2012eb}.
In Refs.~\cite{WalkerLoud:2004hf,Tiburzi:2004rh}, the octet and
decuplet baryon masses were calculated to next-to-next-to-leading
order in heavy baryon chiral perturbation theory and partially
quenched heavy baryon chiral perturbation theory. The
electromagnetic properties of the baryons were calculated in
Refs.~\cite{Wang:2008vb,Camalich:2009uf,Syritsyn:2009mx,Ahuatzin:2010ef,Ledwig:2011cx}.
Since more and more charmed and bottomed baryons were observed
experimentally, there also has been much work on the charmed or
bottomed baryons in the last decade~\cite{Cheng:2006dk,
Detmold:2011bp,Liu:2012uw,Jiang:2014ena,Brown:2014ena,Jiang:2015xqa,Cheng:2015naa,Sun:2016wzh}.
We will mainly investigate the electromagnetic properties of
decuplet baryons in this work.

Historically, the experimental observation of the anomalous magnetic
moment of the nucleon provides the crucial evidence that the nucleon
is not a point particle. In fact, the magnetic moment of the baryon
is an equally important observable as its mass, which encodes
valuable information of its inner structure. In the past several
decades, the magnetic moments of the octet baryons have been
investigated extensively~\cite{Cloet:2014rja,Carrillo-Serrano:2016igi,Zhang:2016qqg}. In fact, their values have been measured\textbf{}
quite precisely~\cite{Agashe:2014kda}. Within the ChPT framework,
the magnetic moments of the octet baryons have been investigated by
many groups
~\cite{Jenkins:1992pi,Meissner:1997hn,Puglia:2000jy,puglia2,Leinweber:1990dv,Savage:2001dy,
Gockeler:2003ay,Arrington:2006zm, Alexandrou:2006ru,Lin:2008mr,Shanahan:2014uka}.

The direct measurement of the magnetic moments of the excited
baryons is difficult because of their short life. However, their
magnetic moments and other electromagnetic form factors of the
short-lived states can be measured from the polarization observables
of the decay products \cite{Aliev:2014foa}, or using the phenomenon
of spin rotation in crystals \cite{Baryshevsky:2016cul}. The study
of the magnetic moments of the nucleon excited states have been
planned at Mainz Microtron (MAMI) facility
\cite{Krusche:2003ik,Kotulla:2002cg,Kotulla:2008zz} and Jefferson
Laboratory \cite{Punjabi:2005wq}. These groups have already realized
the very first effort in measuring the magnetic moments.

The decuplet baryons are the spin-flavor excitations of the octet
baryons. In strong contrast, the present knowledge of the magnetic
moments of the decuplet baryons is rather poor. According to PDG
\cite{Agashe:2014kda}, only the $\Omega^{-}$ magnetic moment is
measured precisely with $\mu_{\Omega^{-}}=(-2.02\pm0.05)\mu_{N}$.
The other members of the decuplet baryons are much more unstable which
renders the experimental measurement of their magnetic moments very
challenging. After huge efforts, the $\Delta^{++}$ and $\Delta^{+}$
magnetic moment were extracted with sizeable uncertainty,
$\mu_{\Delta^{++}}=(5.6\pm1.9) \mu_{N}$ and
$\mu_{\Delta^{+}}=(2.7\pm3.5) \mu_{N}$.

The electromagnetic properties of the decuplet baryons have been
studied in various approaches such as the Skyrme model
~\cite{Adkins:1983ya,Kim:1989qc,Cohen:1986va}, the cloudy-bag model
~\cite{Krivoruchenko:1984xy}, quark models
~\cite{Krivoruchenko:1991pm,Schlumpf:1993rm}, QCD sum rules
~\cite{Aliev:2000rc,Lee:1997jk,Azizi:2008tx,Aliev:2009pd}, chiral
perturbation theory
~\cite{Butler:1993ej,Banerjee:1995wz,Arndt:2003we,Hacker:2006gu,Pascalutsa:2004je,Pascalutsa:2007wb,Geng:2009ys},
lattice
QCD~\cite{Leinweber:1992hy,Cloet:2003jm,Alexandrou:2008bn,Alexandrou:2009hs},
and so on \cite{Segovia:2013uga,Girdhar:2015gsa,Slaughter:2011xs}.
The magnetic dipole and electric quadrupole moments of the decuplet
baryons were computed to the next-to-leading order with chiral
perturbation theory in Ref.~\cite{Butler:1993ej}, where both the
octet and decuplet baryons were included in the chiral loops. In
Ref.~\cite{Banerjee:1995wz}, the Roper contribution to the $\Delta$
magnetic moments was discussed. In
Refs.~\cite{Arndt:2003we,Geng:2009ys}, the electromagnetic
properties of the decuplet baryons were calculated to the
next-to-leading order in the quenched and partially quenched chiral
perturbation theory respectively. In Ref.~\cite{Hacker:2006gu}, the
magnetic dipole moment of the $\Delta(1232)$ was calculated in the
framework of manifestly Lorentz invariant baryon chiral perturbation
theory with the so-called extended on-mass-shell renormalization
scheme. In Refs.~\cite{Pascalutsa:2004je,Pascalutsa:2007wb}, the
authors studied the radiative pion photoproduction on the nucleon
$(\gamma N\rightarrow\pi N \gamma^{\prime})$ in the
$\Delta$-resonance region, with the aim to determine the magnetic
dipole moment (MDM) of the $\Delta^{+}(1232)$. In
Ref.~\cite{Pascalutsa:2006up}, the authors have reviewed the recent
progress in understanding the nature of the $\Delta$-resonance and
its electromagnetic excitation.

In Ref.~\cite{Leinweber:1992hy}, the electromagnetic properties of
the SU(3)-flavor decuplet baryons were examined within a quenched
lattice QCD simulation. The magnetic moments of the $\Delta$ baryons
were extracted from a lattice QCD simulation in
Ref.~\cite{Cloet:2003jm}. Techniques were developed to calculate the
four electromagnetic form factors of the $\Delta$ using lattice QCD
simulation in Refs.~\cite{Alexandrou:2008bn,Alexandrou:2009hs}, with
particular emphasis on the sub-dominant electric quadrupole form
factor that probes the deformation of the $\Delta$. The
electromagnetic form factors of the $\Omega^-$ baryon was studied in
lattice QCD in \cite{Alexandrou:2010}.

Lattice QCD simulation can provide the electromagnetic form factors
from the first principle of QCD. But it usually gives results at
large pion masses. The extrapolated values at the physical pion mass
will be different with different dependence on the pion mass
\cite{Parreno:2016fwu}. With the extrapolating expressions obtained
from ChPT, the electromagnetic form factors of octet baryons
simulated on the lattice are improved obviously
\cite{Shanahan:2014cga,Shanahan:2014uka,Hall:2012pk}. Our work will
also help the extrapolation for the electromagnetic form factors of
the decuplet baryons on the lattice in future.

We investigate the magnetic moments of the decuplet baryons to
$\mathcal{O}(p^{3})$ within the framework of HBChPT at the one-loop
level. The $\mathcal{O}(p^{3})$ results would give some corrections
to the magnetic moments of decuplet baryons as in the case of the
masses and form factors of octet baryons
\cite{Ren:2013dzt,Meissner:1997hn}. Moreover, one cannot judge
whether the chiral expansion up to $\mathcal{O}(p^{2})$ converges or
not without the numerical values of $\mathcal{O}(p^{3})$. We also
discuss the charge radii and  magnetic radii of the decuplet baryons
where the short-distance low energy constant (LEC) is estimated with
the help of the vector meson dominance model and long-range part is
uniquely fixed by the loop corrections.

We explicitly consider both the octet and decuplet intermediate
states in the loop calculation because the mass splitting between
the octet and decuplet baryons is small. Moreover, the decuplet
baryons generally couple to the octet baryons strongly. For example,
the $\Delta$ resonance couples to the $N\pi$ channel very strongly.
We use the dimensional regularization and modified minimal
subtraction scheme to deal with the divergences from the loop
corrections.

We will calculate the charge (E0), electro quadrupole (E2), magnetic
dipole (M1) and magnetic octupole (M3) form factors of the decuplet
baryons in the framework of HBChPT. In the limit $q^2=0$, we extract
the magnetic moments of the decuplet baryons. Since the experimental
measurement of the electro quadrupole and magnetic octupole form
factors of the decuplet baryons will be extremely difficult in the
coming future, we move the calculation and discussions of these two
form factors to the Appendix. In the text, we focus on the
calculation of the charge and magnetic form factors.

This paper is organized as follows. In Section \ref{Sec3}, we
discuss the electromagnetic form factors of the spin-$\frac{3}{2}$
particles. We introduce the effective chiral Lagrangians of the
decuplet baryon in Section \ref{Sec2}. In Section \ref{secFormalism}, we
calculate the multipole form factors of the decuplet baryons order by
order. We estimate the low-energy constants in Section \ref{SecEstLEC}.
We present our numerical results in Section \ref{Sec6} and
conclude in Section \ref{Sec7}. We collect some useful formulae and
the coefficients of the loop corrections in the appendix.

%%%%%%%%%%%%%%%%%%%%%%%%%%%%%%%%%%%%%%%%%%%%
\section{The electromagnetic form factors of the decuplet baryons}\label{Sec3}
%%%%%%%%%%%%%%%%%%%%%%%%%%%%%%%%%%%%%%%%%%%%

\subsection{The multipole form factors}

When the electromagnetic current is sandwiched between two decuplet
baryon states, one can write down the general matrix elements which
satisfy the gauge invariance, parity conservation and time reversal
invariance ~\cite{Nozawa:1990gt}:
\begin{equation}
<T(p^{\prime})|J_{\mu}|T(p)>=-\bar{u}^{\rho}(p^{\prime})O_{\rho\mu\sigma}(p^{\prime},p)u^{\sigma}(p),
\end{equation}
where
\begin{equation}
O_{\rho\mu\sigma}(p^{\prime},p)=g_{\rho\sigma}(A_{1}\gamma_{\mu}+\frac{A_{2}}{2M_{T}}P_{\mu})
+\frac{q_{\rho}q_{\sigma}}{(2M_{T})^{2}}(C_{1}\gamma_{\mu}+\frac{C_{2}}{2M_{T}}P_{\mu}).
\label{eq_new_current}
\end{equation}
where $p$ and $p'$ are the momenta of decuplet baryons. In the above equations, $P=p^{\prime}+p$, $q=p^{\prime}-p$, $M_{T}$ is decuplet-baryon mass, and
$u_{\rho}(p)$ is the Rarita-Schwinger spinor for an on-shell heavy
baryon satisfying $p^{\rho}u_{\rho}(p)=0$ and
$\gamma^{\rho}u_{\rho}(p)=0$. $A_{1,2}$ and $C_{1,2}$ are real
functions of $q^{2}$. In literature, there exists another definition
of the tensor $O_{\rho\mu\sigma}(p^{\prime},p)$ ~\cite{H. Arenhovel}
\begin{eqnarray}
\begin{split}
O_{\rho\mu\sigma}(p^{\prime},p) & = & g_{\rho\sigma}(a_{1}\gamma_{\mu}+a_{2}P_{\mu})+a_3(q_{\rho}g_{\mu\sigma}-g_{\rho\mu}q_{\sigma})\\
 &  & +q_{\rho}q_{\sigma}(c_{1}\gamma_{\mu}+c_{2}P_{\mu})+ic_3\gamma_{5}\epsilon_{\rho\mu\sigma\lambda}q^{\lambda},
\label{eq_old_current}
\end{split}
\end{eqnarray}
where $a_{i}$ and $c_{i}$ are real functions of $q^{2}$.
$\epsilon_{\rho\mu\sigma\lambda}$ is the totally antisymmetric
rank-4-tensor with $\epsilon_{0123}=1$. However, the expression in
Eq.~(\ref{eq_old_current}) contains two additional terms ($b$ term and
$d$ term) which are not linearly independent of the other terms. For
example, the tensor structure
$(q_{\rho}g_{\mu\sigma}-g_{\rho\mu}q_{\sigma})$ is not dependent if
both the initial and final decuplet baryons are on-shell.
\begin{eqnarray}
\bar{u}^{\rho}(p^{\prime})\left(q_{\rho}g_{\mu\sigma}-g_{\rho\mu}q_{\sigma}\right)
u^{\sigma}(p) = \bar{u}^{\rho}(p^{\prime})\left[
2M_{T}(1-\frac{q^{2}}{4M_{T}^{2}})g_{\rho\sigma}\gamma_{\mu}
-g_{\rho\sigma}P_{\mu}+\frac{1}{M_{T}}q_{\rho}q_{\sigma}\gamma_{\mu}
\right] u^{\sigma}(p).
\end{eqnarray}
In the following, we shall use Eq.~(\ref{eq_new_current}) to define
the charge (E0), electro quadrupole (E2), magnetic-dipole (M1) and
magnetic octupole (M3) multipole form factors of the decuplet
baryons
\begin{equation}
\begin{cases} \displaystyle
G_{E0}(q^{2})  =  (1+\frac{2}{3}\tau)[A_{1}+(1+\tau)A_{2}]-\frac{1}{3}\tau(1+\tau)[C_{1}+(1+\tau)C_{2}],\\ \displaystyle
G_{E2}(q^{2})  =  [A_{1}+(1+\tau)A_{2}]-\frac{1}{2}(1+\tau)[C_{1}+(1+\tau)C_{2}],\\ \displaystyle
G_{M1}(q^{2})  =  (1+\frac{4}{5}\tau)A_{1}-\frac{2}{5}\tau(1+\tau)C_{1},\\ \displaystyle
G_{M3}(q^{2})  =  A_{1}-\frac{1}{2}(1+\tau)C_{1},
\end{cases}
\end{equation}
where $\tau=-\frac{q^{2}}{(2M_{T})^{2}}$.

With $q^2=0$, we obtain the charge, electro quadrupole moment,
magnetic moment, magnetic octupole moment and charge radii of the
decuplet baryons etc.
\begin{equation}
\begin{cases} \displaystyle
G_{E0}(0)  =  A_{1}+A_{2},\\ \displaystyle
G_{E2}(0)  =  A_{1}+A_{2}-\frac{1}{2}(C_{1}+C_{2}),\\ \displaystyle
G_{M1}(0)  =  A_{1},\\ \displaystyle
G_{M3}(0)  =  A_{1}-\frac{1}{2}C_{1}\\ \displaystyle
\langle{r^{2}}\rangle=6\frac{dG_{E0}(q^{2})}{dq^{2}}\mid_{q^{2}=0}.
\end{cases}
\end{equation}

\subsection{The form factors in the non-relativistic limit}

In the heavy baryon limit, the baryon field $B$ can be decomposed
into the large component $\mathcal{N}$ and the small component
$\mathcal{H}$.
\begin{equation}
B=e^{-iM_{B}v\cdot x}(\mathcal{N}+\mathcal{H}),
\end{equation}
\begin{equation}
\mathcal{N}=e^{iM_{B}v\cdot x}\frac{1+v\hspace{-0.5em}/}{2}B,~
\mathcal{H}=e^{iM_{B}v\cdot x}\frac{1-v\hspace{-0.5em}/}{2}B,
\end{equation}
where $M_{B}$ is the octet-baryon mass, $v_{\mu}=(1,\vec{0})$ is the velocity of the baryon. For the
decuplet baryon, the large component is denoted as
$\mathcal{T}_{\mu}$. Now the decuplet matrix elements of the
electromagnetic current $J_{\mu}$ can be parameterized as
\begin{equation}
<\mathcal T(p^{\prime})|J_{\mu}|\mathcal T(p)>=-\bar{u}^{\rho}(p^{\prime})\mathcal O_{\rho\mu\sigma}(p^{\prime},p)u^{\sigma}(p).
\end{equation}
The tensor $\mathcal O_{\rho\mu\sigma}$ can be parameterized in terms of four
Lorentz invariant form factors.
\begin{eqnarray}
\begin{split}
\mathcal O_{\rho\mu\sigma}(p^{\prime},p)=g_{\rho\sigma}\left[v_{\mu}F_{1}(q^{2})+\frac{[S_{\mu},S_{\alpha}]}
{M_{T}}q^{\alpha}F_{2}(q^{2})\right]
+\frac{q^{\rho}q^{\sigma}}{(2M_{T})^{2}}\left[v_{\mu}F_{3}(q^{2})+\frac{[S_{\mu},S_{\alpha}]}
{M_{T}}q^{\alpha}F_{4}(q^{2})\right]. \label{eq_newnew_current}
\end{split}
\end{eqnarray}
The multipole form factors are
\begin{eqnarray}
\begin{cases} \displaystyle
G_{E0}(q^{2})  =  (1+\frac{2}{3}\tau)[F_{1}+\tau(F_{1}-F_{2})]-\frac{1}{3}\tau(1+\tau)[F_{3}+\tau(F_{3}-F_{4})], \\ \displaystyle
G_{E2}(q^{2})  =  [F_{1}+\tau(F_{1}-F_{2})]-\frac{1}{2}(1+\tau)[F_{3}+\tau(F_{3}-F_{4})], \\ \displaystyle
G_{M1}(q^{2})  =  (1+\frac{4}{5}\tau)F_{2}-\frac{2}{5}\tau(1+\tau)F_{4},\\ \displaystyle
G_{M3}(q^{2})  =  F_{2}-\frac{1}{2}(1+\tau)F_{4}.
\end{cases}
\label{eq_formfactor}
\end{eqnarray}
Accordingly, the multipole form factors at $q^2=0$ lead to the
charge ($Q$), the magnetic dipole moment ($\mu$), the electric
quadrupole moment ($\mathbb{Q}$), and the magnetic octupole moment
($\mathcal{O}$):
\begin{equation}
\begin{cases}
Q=G_{E0}(0)  =  F_1,\\  \displaystyle
\mathbb{Q}=\frac{e}{M_{T}^2}G_{E2}(0)  =  \frac{e}{M_{T}^2}(F_1-\frac{1}{2}F_3),\\  \displaystyle
\mu=\frac{e}{2M_{T}}G_{M1}(0)  =  \frac{e}{2M_{T}}F_2,\\  \displaystyle
\mathcal{O}=\frac{e}{2M_{T}^3}G_{M3}(0)  =  \frac{e}{2M_{T}^3}(F_2-\frac{1}{2}F_4)\\  \displaystyle
\langle{r_{E}^{2}}\rangle=6\frac{dG_{E0}(q^{2})}{dq^{2}}\mid_{q^{2}=0}.
\end{cases}
\label{eqDefMoments}
\end{equation}

%%%%%%%%%%%%%%%%%%%%%%%%%%%%%%%%%%%%%%%%%%%%%%%
\section{Chiral Lagrangians}\label{Sec2}
%%%%%%%%%%%%%%%%%%%%%%%%%%%%%%%%%%%%%%%%%%%%%%%

%%%%%%%%%%%%%%%%%%%%%
\subsection{The strong interaction chiral Lagrangians}
%%%%%%%%%%%%%%%%%%%%%

The pseudoscalar meson fields are introduced as follows,
\begin{equation}
\phi=\left(\begin{array}{ccc}
\pi^{0}+\frac{1}{\sqrt{3}}\eta & \sqrt{2}\pi^{+} & \sqrt{2}K^{+}\\
\sqrt{2}\pi^{-} & -\pi^{0}+\frac{1}{\sqrt{3}}\eta & \sqrt{2}K^{0}\\
\sqrt{2}K^{-} & \sqrt{2}\bar{K}^{0} & -\frac{2}{\sqrt{3}}\eta
\end{array}\right).
\end{equation}
In the framework of ChPT, the chiral connection and axial vector
field are defined as~\cite{Scherer:2002tk,Bernard:1995dp},
\begin{equation}
\Gamma_{\mu}=\frac{1}{2}\left[u^{\dagger}(\partial_{\mu}-ir_{\mu})u+u(\partial_{\mu}-il_{\mu})u^{\dagger}\right],
\end{equation}
\begin{equation}
u_{\mu}\equiv\frac{1}{2}i\left[u^{\dagger}(\partial_{\mu}-ir_{\mu})u-u(\partial_{\mu}-il_{\mu})u^{\dagger}\right],
\end{equation}
where
\begin{equation}
u^{2}=\mathit{U}=\exp(i\phi/f_{0}).
\end{equation}
$f_0$ is the decay constant of the pseudoscalar meson in the chiral
limit. The experimental value of the pion decay constant
$f_{\pi}\approx$ 92.4 MeV while $f_{K}\approx$ 113 MeV, $f_{\eta}\approx$ 116 MeV.

The lowest order ($\mathcal{O}(p^{2})$) pure meson Lagrangian is
\begin{equation}
\mathcal{L}_{\pi\pi}^{(2)}=\frac{f_{0}^{2}}{4}{\rm
Tr}[\nabla_{\mu}U(\nabla^{\mu}U)^{\dagger}] \label{Eq:meson1},
\end{equation}
where
\begin{equation}
\nabla_{\mu}U=\partial_{\mu}U-ir_{\mu}U+iUl_{\mu}.
\end{equation}
For the electromagnetic interaction,
\begin{equation}
r_{\mu}=l_{\mu}=-eQA_{\mu},Q=\rm{diag}(\frac{2}{3},-\frac{1}{3},-\frac{1}{3}).
\end{equation}
The spin-$\frac{1}{2}$ octet field reads
\begin{equation}
B=\left(\begin{array}{ccc}
\frac{1}{\sqrt{2}}\Sigma^{0}+\frac{1}{\sqrt{6}}\Lambda & \Sigma^{+} & p\\
\Sigma^{-} & -\frac{1}{\sqrt{2}}\Sigma^{0}+\frac{1}{\sqrt{6}}\Lambda & n\\
\Xi^{-} & \Xi^{0} & -\frac{2}{\sqrt{6}}\Lambda
\end{array}\right).
\end{equation}
For the spin-$\frac{3}{2}$ decuplet field, we adopt the
Rarita-Schwinger field $T^{\mu}\equiv
{T^{\mu}}^{abc}$~\cite{Rarita:1941mf}:
\begin{eqnarray}
&&T^{111}=\Delta^{++},T^{112}=\frac{1}{\sqrt{3}}\Delta^{+},T^{122}=\frac{1}{\sqrt{3}}\Delta^{0},T^{222}=\Delta^{-},T^{113}=\frac{1}{\sqrt{3}}\Sigma^{*+},\nonumber\\
&&T^{123}=\frac{1}{\sqrt{6}}\Sigma^{*0},T^{223}=\frac{1}{\sqrt{3}}\Sigma^{*-},T^{133}=\frac{1}{\sqrt{3}}\Xi^{*0},T^{233}=\frac{1}{\sqrt{3}}\Xi^{*-},T^{333}=\Omega^{-}.
\end{eqnarray}
The leading order pseudoscalar meson and baryon interaction
Lagrangians read~\cite{Rarita:1941mf,Jenkins:1992pi}
\begin{eqnarray}
\hat{\mathcal{L}}_{0}^{(1)}&=&{\rm Tr}[\bar{B}(iD\hspace{-0.6em}/-M_{B})B] \nonumber\\
&&+{\rm Tr}{\bar
T^{\mu}[-g_{\mu\nu}(iD\hspace{-0.6em}/-M_{T})+i(\gamma_{\mu}
D_{\mu}+\gamma_{\nu}D_{\mu})-\gamma_{\mu}(iD\hspace{-0.6em}/+M_{T})\gamma_{\nu}]T^{\nu}},
\label{Eq:baryon01}\\
\hat{\mathcal{L}}_{\rm int}^{(1)}&=&\mathcal{C}[{\rm Tr}(\bar
T^{\mu}u_{\mu}B)+{\rm Tr}(\bar B
u_{\mu}T^{\mu})]+\mathcal{H}{\rm Tr}(
\bar T^{\mu}g_{\mu\nu}u\hspace{-0.5em}/
\gamma_{5}T^{\nu}) ,\label{Eq:baryon02}
\end{eqnarray}
where $M_{B}$ is octet-baryon mass, $M_{T}$ is decuplet-baryon mass,
\begin{eqnarray}
D_{\mu}B&=&\partial_{\mu}B+[\Gamma_{\mu},B], \nonumber\\
D^{\nu}(T^{\mu})_{abc}&=&\partial^{\nu}(T^{\mu})_{abc}+(\Gamma^{\nu})^{d}_{a}(T^{\mu})_{dbc}+(\Gamma^{\nu})^{d}_{b}(T^{\mu})_{adc}+(\Gamma^{\nu})^{d}_{c}(T^{\mu})_{abd}.
\end{eqnarray}
We also need the second order pseudoscalar meson and decuplet baryon
interaction Lagrangian.
\begin{equation}
\hat{\mathcal{L}}_{\rm int}^{(2)}=\frac{ig_{2}}{4M_{B}} {\rm Tr}(
g_{\rho\sigma}\bar{T}^{\rho}[u_{\mu},u_{\nu}]\sigma^{\mu\nu}T^{\sigma}),
\label{Eq:baryon03}
\end{equation}
where the superscript denotes the chiral order and $g_2$ is the
coupling constant.

In the framework of HBChPT, the baryon field $B$ is decomposed into
the large component $\mathcal{N}$ and the small component
$\mathcal{H}$. We denote the large component of the decuplet baryon
as $\mathcal{T}_{\mu}$. The leading order nonrelativistic
pseudoscalar meson and baryon Lagrangians read~\cite{Jenkins:1992pi}
\begin{equation}
\mathcal{L}_{0}^{(1)}={\rm Tr}[\bar{\mathcal{N}}(iv\cdot
D-\delta)\mathcal{N}]-i\bar{\mathcal{T}}^{\mu}(v\cdot
D)\mathcal{T}_{\mu}, \label{Eq:baryon1}
\end{equation}
\begin{equation}
\mathcal{L}_{\rm int}^{(1)}=\mathcal{C}(\bar{\mathcal{T}}^{\mu}u_{\mu}\mathcal{N}+\bar{\mathcal{N}}u_{\mu}\mathcal{T}^{\mu})+2\mathcal{H}
\bar{\mathcal{T}}^{\mu}S^{\nu}u_{\nu}\mathcal{T}_{\mu},
\label{Eq:baryon2}
\end{equation}
where $\mathcal{L}_{0}^{(1)}$ and $\mathcal{L}_{\rm int}^{(1)}$ are the
free and interaction parts respectively. $S_{\mu}$ is the covariant
spin-operator. $\delta=M_{B}-M_{T}$ is the octet and decuplet baryon
mass splitting. In the isospin symmetry limit, $\delta = -0.2937$
GeV. We do not consider the mass difference among different decuplet
baryons. The $\phi\mathcal{N}\mathcal{T}$ coupling
$\mathcal{C}=-1.2\pm0.1$ while the $\phi\mathcal{T}\mathcal{T}$ coupling
$\mathcal{H}=-2.2\pm0.6$~\cite{Butler:1992pn}. For the pseudoscalar mesons  masses, we use
$m_{\pi}=0.140$ GeV, $m_{K}=0.494$ GeV, and $m_{\eta}=0.550$ GeV. We use the averaged masses for the octet and decuplet baryons, and $M_{B}=1.158$ GeV, $M_{T}=1.452$ GeV.

The second order nonrelativistic pseudoscalar meson and baryon
Lagrangian reads,
\begin{equation}
\mathcal{L}_{\rm int}^{(2)}=\frac{g_{2}}{2M_{B}}{\rm Tr}(
g_{\rho\sigma}\bar{\mathcal{T}}^{\rho}[S^\mu,S^\nu][u_\mu,u_\nu]\mathcal{T}^{\sigma}),
\label{Eq:TTUU}
\end{equation}
where $g_{2}$ is the $\phi\phi\mathcal{T}\mathcal{T}$ coupling
constant to be determined. In fact, there exist several
$\phi\phi\mathcal{T}\mathcal{T}$ interaction terms with other
Lorentz structures. However, these additional terms do not
contribute to the present investigations of the electromagnetic form
factors of the decuplet baryons. So we omit them and keep the $g_2$
term only.

%%%%%%%%%%%%%%%%%%%%%%%%%%%%%
\subsection{The electromagnetic chiral Lagrangians at $\mathcal{O}(p^{2})$}
%%%%%%%%%%%%%%%%%%%%%%%%%%%%%

The lowest order $\mathcal{O}(p^{2})$ Lagrangian contributes to the
magnetic moments and magnetic dipole form factors of the decuplet
baryons at the tree level~\cite{Jenkins:1992pi}
\begin{equation}
\mathcal{L}_{\mu_{\mathcal T}}^{(2)}=\frac{-i}{2M_{B}}{\rm
Tr}\bar{\mathcal{T}}^{\mu} (b-b_{q^2}\partial^2)F_{\mu\nu}^{+}
\mathcal{T}^{\nu}, \label{Eq:baryon3}
\end{equation}
where the coefficients $b$ and $b_{q^2}$ are new LECs which
contribute to the magnetic moments and magnetic radii of the
decuplet baryons at the tree-level respectively. The chirally
covariant QED field strength tensor $F_{\mu\nu}^{\pm}$ is defined as
\begin{eqnarray} \nonumber
F_{\mu\nu}^{\pm} & = & u^{\dagger}F_{\mu\nu}^{R}u\pm
uF_{\mu\nu}^{L}u^{\dagger},\\
F_{\mu\nu}^{R} & = &
\partial_{\mu}r_{\nu}-\partial_{\nu}r_{\mu}-i[r_{\mu},r_{\nu}],\\
F_{\mu\nu}^{L} & = &
\partial_{\mu}l_{\nu}-\partial_{\nu}l_{\mu}-i[l_{\mu},l_{\nu}],
\end{eqnarray}
where $r_{\mu}=l_{\mu}=-eQA_{\mu}$. The operator $F_{\mu\nu}^{\pm}$
transforms as the adjoint representation. Recall that the direct
product $10\otimes\bar{10}=1\oplus8\oplus27\oplus64$ contains only
one adjoint representation. Therefore, there is only one independent
interaction term in the $\mathcal{O}(p^{2})$ Lagrangians for the
magnetic moments of the decuplet baryons.

The lowest order Lagrangians which contribute to the magnetic
moments of the octet baryons at the tree level are~\cite{Meissner:1997hn},
\begin{equation}
\mathcal{L}_{\mu_{\mathcal N}}^{(2)}=b_F\frac{-i}{4M_{B}}{\rm
Tr}\bar{\mathcal{N}}[S^{\mu},S^{\nu}] [F_{\mu\nu}^{+},
\mathcal{N}]+b_D\frac{-i}{4M_{B}}{\rm
Tr}\bar{\mathcal{N}}[S^{\mu},S^{\nu}]
\{F_{\mu\nu}^{+},\mathcal{N}\}. \label{Eq:baryonoctet}
\end{equation}

The lowest order Lagrangians which contribute to the decuplet-octet
transition magnetic moments at the tree level are
\begin{equation}
\mathcal{L}_{\mu_{\mathcal T\mathcal N}}^{(2)}=b_2\frac{-i}{2M_{B}}{\rm
Tr}\bar{\mathcal{T}}^{\mu} F_{\mu\nu}^{+}S^{\nu}\mathcal{N}
+b_3\frac{-i}{2M_{B}}{\rm Tr}\bar{\mathcal{T}}^{\mu}
F_{\mu\nu}^{+}D^{\nu}\mathcal{N}+{\rm H.c.},\label{Eq:baryon_trans}
\end{equation}
where $b_2=2.4$ is estimated with the help of quark model. The $b_3$
term does not contribute to the magnetic moments of the decuplet
baryons.

%%%%%%%%%%%%%%%%%%%%%%%%
\subsection{The higher order electromagnetic chiral Lagrangians }
%%%%%%%%%%%%%%%%%%%%%%%%

We also need the $\mathcal{O}(p^{3})$ Lagrangian which contributes
to the short-distance part of the charge radii
\begin{equation}
\mathcal{L}_{r}^{(3)}=\frac{-c_{r}}{4M_T^{2}}{\rm
Tr}\bar{\mathcal{T}}^{\rho}\mathcal{T}_{\rho}v^{\mu}
\partial^{\nu}F_{\mu\nu}^{+}.
\label{Eq:chargeradii}
\end{equation}

The $\mathcal{O}(p^{3})$ Lagrangian which contributes to the electro
quadrupole moments and its radii at the tree
level~\cite{Butler:1993ej,Arndt:2003we} reads
\begin{equation}
\mathcal{L}_{\mathbb{Q}}^{(3)}=\frac{c_{\mathbb{Q}}}{4M_T^{2}} {\rm
Tr}\bar{\mathcal{T}}^{\{\rho}\mathcal{T}^{\sigma\}}
v^{\mu}\partial_{\rho}F_{\mu\sigma}^{+}
,\label{Eq:electro quadrupole}
\end{equation}
where
$\bar{\mathcal{T}}^{\{\rho}\mathcal{T}^{\sigma\}}=\bar{\mathcal{T}}^{\rho}\mathcal{T}^{\sigma}
+\bar{\mathcal{T}}^{\sigma}\mathcal{T}^{\rho}-\frac{1}{2}g^{\rho\sigma}\bar{\mathcal{T}}^{\alpha}\mathcal{T}_{\alpha}$.

To calculate the magnetic moments to $\mathcal{O}(p^{4})$, we also need the $\mathcal{O}(p^{4})$ electromagnetic chiral
Lagrangians at the tree level. Recall that
\begin{eqnarray}
10\otimes\bar{10}&=&1\oplus8\oplus27\oplus64,\\
8\otimes8&=&1\oplus8_{1}\oplus8_{2}\oplus10\oplus\bar{10}\oplus27.
\end{eqnarray}
Both $F_{\mu\nu}^{\pm}$ and $\chi^{+}$ transform as the adjoint
representation. When the product $F_{\mu\nu}^{\pm} \chi^{+}$ belongs
to the $1, 8_1, 8_2$ and $27$ flavor representation, we can write
down the chirally invariant $\mathcal{O}(p^{4})$ electromagnetic
Lagrangians. Therefore, it seems there should be four independent
interaction terms in the $\mathcal{O}(p^{4})$ chiral Lagrangians.
However, it only contains three independent terms after considering
C parity,
\begin{eqnarray}
\mathcal{L}_{\mu}^{(4)}&=&d_1\frac{-i}{2M_{B}}{\rm Tr}(\bar{\mathcal{T}}^{\mu}\mathcal{T}^{\nu}){\rm Tr}(\chi^{+}F_{\mu\nu}^{+})\nonumber\\
&&+d_2\frac{-i}{2M_{B}}{\rm
Tr}(\bar{\mathcal{T}}^{\mu}_{ijk}({F_{\mu\nu a}^{+~i}}{\chi^{+k}_{~l}}){\mathcal{T}^{\nu}}^{ajl})
+d_3\frac{-i}{2M_{B}}{\rm
Tr}(\bar{\mathcal{T}}^{\mu}_{ijk}({F_{\mu\nu}^{+}}{\chi^{+}})^{i}_{l}{\mathcal{T}^{\nu}}^{ljk}).
\label{Eq:baryonp40}
\end{eqnarray}
where $\chi^{+}$=diag(0,0,1) at the leading order and the factor
$m_{s}$ has been absorbed in the LECs $d_{1,2,3}$.

There is one more term which contributes to the decuplet magnetic
moments,
\begin{equation}
\mathcal{L'}_{\mu}^{(4)}=b^{\prime}\frac{-i}{2M_{B}}{\rm
Tr}(\bar{\mathcal{T}}^{\mu} F_{\mu\nu}^{+} \mathcal{T}^{\nu}){\rm
Tr}(\chi^{+}). \label{Eq:baryon_4order}
\end{equation}
However, its contribution can be absorbed through the renomalization
of the LEC $b$, i.e.
\begin{eqnarray}
b\rightarrow b+{\rm Tr}(\chi^{+})b^{\prime}.
\end{eqnarray}

The $\mathcal{O}(p^{4})$ Lagrangian which contributes to the
magnetic octupole moments and its radii at the tree level is
constructed as
\begin{equation}
\mathcal{L}_{\mathcal{O}}^{(4)}=\frac{-d_{\mathcal{O}}}{8M_T^{3}}{\rm
Tr}\bar{\mathcal{T}}^
{\{\rho}\mathcal{T}^{\sigma\}}\sigma^{\mu\nu}\partial_{\nu}\partial_{\rho}F_{\mu\sigma}^{+}.
\label{Eq:magnetic octupole}
\end{equation}

%%%%%%%%%%%%%%%%%%%%%%%%%%%%%%%%%%%%%%%%%%%
\section{Formalism up to one-loop level}\label{secFormalism}
%%%%%%%%%%%%%%%%%%%%%%%%%%%%%%%%%%%%%%%%%%%
We apply the standard power counting scheme of HBChPT. The chiral
order $D_{\chi}$ of a given diagram is given by~\cite{Ecker:1994gg}
\begin{equation}
D_{\chi}=4N_{L}-2I_{M}-I_{B}+\sum_{n}nN_{n},
\label{Eq:Power counting}
\end{equation}
where $N_{L}$ is the number of loops, $I_{M}$ is the number of
internal pion lines, $I_{B}$ is the number of internal octet or
decuplet nucleon lines and $N_{n}$ is the vertices from the $n$th
order Lagrangians. As an example, we consider the one-loop diagram a
in Fig.~\ref{fig:allloop}. First of all, the number of independent
loops $N_{L}=1$, the number of internal pion lines $I_{M}=1$, the
number of internal octet or decuplet nucleon lines  $I_{B}=2$. For
$N_{1}=2$, and $N_{2}=1$ we obtain $D_{\chi}=4-2-2+2+2=4$.

We use Eq. (\ref{Eq:Power counting}) to count the chiral order $D_\chi$ of the matrix element of the current, $e \mathcal
O_{\rho\mu\sigma}$. We count the unit charge $e$ as $\mathcal
O(p^1)$. The chiral orders of $F_1$, $F_2$, $F_3$, and $F_4$ are $(D_\chi-1)$, $(D_\chi-2)$, $(D_\chi-3)$, and $(D_\chi-4)$, respectively, since
\begin{equation}
e \mathcal O_{\rho\mu\sigma} \sim e p^0 F_1+e p^1 F_2+e p^2 F_3+e p^3 F_4.
\end{equation}
The chiral order of magnetic dipole moments $\mu$ is $(D_\chi-1)$ based on Eq. (\ref{eqDefMoments}).

%%%%%%%%%%%%%%%%%%%%%%%%%%%%%%%%%%%%%%%%%%%
\subsection{The magnetic moments}
\begin{figure}
\begin{minipage}[t]{0.4\linewidth}
\centering
\includegraphics[width=0.5\hsize]{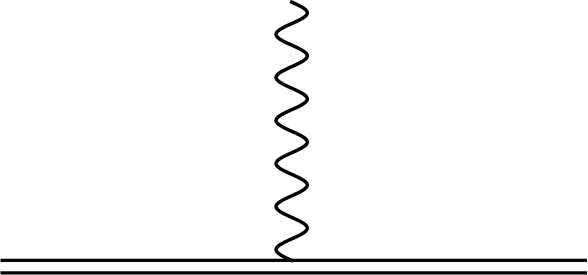}
%\caption{The $\mathcal{O}(p^{2})$ tree level diagram}
%\label{fig:tree2}
\end{minipage}%
\begin{minipage}[t]{0.4\linewidth}
\centering
\includegraphics[width=0.5\hsize]{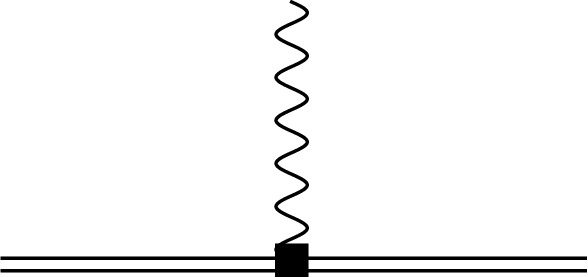}
%\caption{The $\mathcal{O}(p^{4})$ tree level diagram}
%\label{fig:tree4}
\end{minipage}
\caption{The $\mathcal{O}(p^{2})$ and $\mathcal{O}(p^{4})$ tree level diagram. The left dot and the right black square represent second- and fourth-order
couplings respectively.}
\label{fig:tree}
\end{figure}

%%%%%%%%%%%%%%%%%%%%%%%%%%%%%%%%%%%%%%%%%%%
Throughout this work, we assume the exact isospin symmetry with
$m_{u}=m_{d}$. The tree-level Lagrangians in Eqs.
~(\ref{Eq:baryon3}),(\ref{Eq:baryonp40}) contribute to the decuplet
magnetic moments at $\mathcal{O}(p^{1})$ and $\mathcal{O}(p^{3})$ as
shown in Fig.~\ref{fig:tree}. The
Clebsch-Gordan coefficients for the various decuplet states are
collected in Table~\ref{Magnetic moments}. All
decuplet magnetic moments are given in terms of the LECs $b$,
$d_{1}$, $d_{2}$ and $d_{3}$. There exist several interesting
relations,
\begin{eqnarray}
%&&\mu^{\rm tree}_{\Delta^{++}}-\mu^{\rm tree}_{\Delta^{+}}-\mu^{\rm tree}_{\Delta^{0}}+\mu^{\rm tree}_{\Delta^{-}}=0,\nonumber\\
%&&\mu^{\rm tree}_{\Delta^{++}}-\mu^{\rm tree}_{\Delta^{-}}-3(\mu^{\rm tree}_{\Delta^{+}}-\mu^{\rm tree}_{\Delta^{0}})=0,\nonumber\\
&&2\mu^{\rm tree}_{\Sigma^{0}}=\mu^{\rm tree}_{\Sigma^{+}}+\mu^{\rm tree}_{\Sigma^{-}},\nonumber\\
&&2\mu^{\rm tree}_{\Sigma^{0}}=\mu^{\rm tree}_{\Delta^{0}}+\mu^{\rm tree}_{\Xi^{0}},\nonumber\\
&&2\mu^{\rm tree}_{\Sigma^{-}}=\mu^{\rm tree}_{\Xi^{-}}+\mu^{\rm tree}_{\Delta^{-}},\nonumber\\
&&\mu^{\rm tree}_{\Omega^{-}}=\mu^{\rm tree}_{\Sigma^{-}}+\mu^{\rm tree}_{\Xi^{-}}-\mu^{\rm tree}_{\Delta^{-}}.
\end{eqnarray}

\begin{figure}[tbh]
\centering
\includegraphics[width=0.9\hsize]{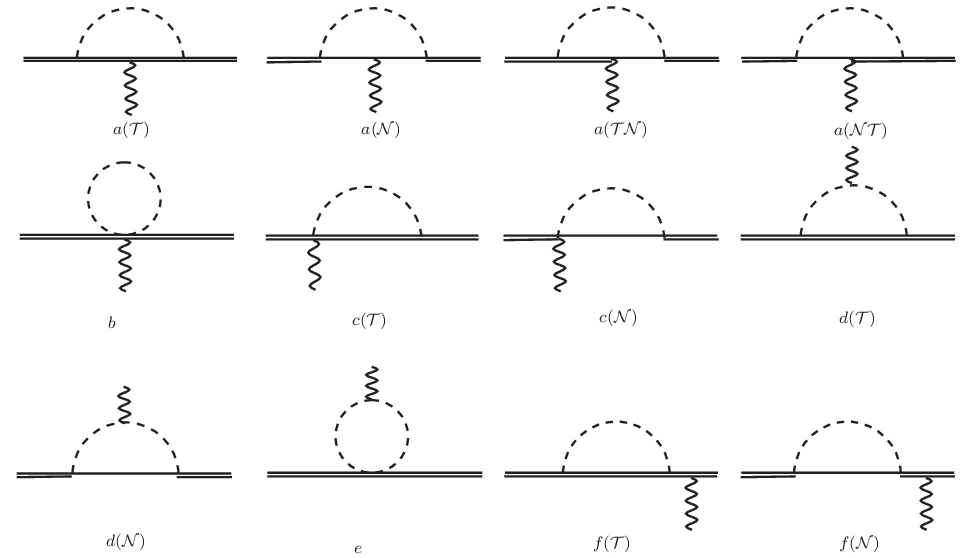}
\caption{The one-loop diagrams where the decuplet (octet) baryon is
denoted by the double (single) solid line. The dashed and wiggly
lines represent the pseudoscalar meson and photon respectively. For
the wave function renormalization in diagram f, only representative
graphs are shown. We do not also list the adjoint graphs for diagram
c.}\label{fig:allloop}

\end{figure}

There are twelve Feynman diagrams at one-loop level as shown in
Fig.~\ref{fig:allloop} and we divide them into six types (a-f)
according to the structure. All the vertices in these diagrams come
from
Eqs.~(\ref{Eq:meson1}),(\ref{Eq:baryon1}-\ref{Eq:baryon_trans}). In
diagram a, the meson vertex is from the strong interaction terms in
Eq.~(\ref{Eq:baryon2}) while the photon vertex from the
$\mathcal{O}(p^{2})$ tree level magnetic moment interaction in
Eqs.~(\ref{Eq:baryon3}),(\ref{Eq:baryonoctet}),(\ref{Eq:baryon_trans}).
In diagram b, the photon-meson-baryon vertex is also from the
$\mathcal{O}(p^{2})$ tree level magnetic moment interaction in
Eq.~(\ref{Eq:baryon3}). In diagram c, the two vertices are from the
strong interaction and seagull terms respectively. In diagram d, the
meson vertex is from the strong interaction terms while the photon
vertex is from the meson photon interaction term in
Eq.~(\ref{Eq:meson1}). In diagram e, the meson-baryon vertex is from
the second order pseudoscalar meson and baryon Lagrangian in
Eq.~(\ref{Eq:TTUU}) while the photon vertex is also from the meson
photon interaction term. In diagram f, the meson vertex is from the
strong interaction terms while the photon vertex from the
$\mathcal{O}(p^{2})$ tree level magnetic moment interaction.

The diagrams a, b, e and f contribute to the tensor $e \mathcal O_{\rho\mu\sigma}$
at $\mathcal{O}(p^{4})$ while the diagram d contributes at
$\mathcal{O}(p^{3})$. The diagram c vanishes in the heavy baryon
mass limit. If the intermediate baryon is a decuplet (or octet)
state, the amplitude of the diagram c is denoted as $J_{c(\mathcal T)}$ (or
$J_{c(\mathcal N)}$). We have
\begin{eqnarray}
 J_{c(\mathcal T)} & \propto & \int\frac{d^{d}l}{(2\pi)^{d}}\frac{i}{l^{2}-m^{2}_\phi+i\epsilon}\frac{g_{\mathcal{H}}(S\cdot l)}{f_{0}}
 \frac{-iP_{\rho\sigma}^{3/2}}{v\cdot l+i\epsilon}S_\mu\protect\\
 \nonumber
 & \propto & S\cdot v=0,\\
  J_{c(\mathcal N)} & \propto & \int\frac{d^{d}l}{(2\pi)^{d}}\frac{i}{l^{2}-m^{2}_\phi+i\epsilon}\frac{g_{\mathcal{H}} l_\sigma}{f_{0}}
 \frac{i}{v\cdot l- \omega +i\epsilon}g_{\mu\rho}\protect\\  \nonumber
 & \propto & g_{\mu\rho}v_{\sigma},
 \end{eqnarray}
where $P_{\rho\sigma}^{3/2}$ is the non-relativistic spin-$\frac32$ projector. $J_{c(\mathcal T)}$ vanishes, and $J_{c(\mathcal N)}$ also vanishes since
$v_{\sigma}u^{\sigma}=0$. In other words, this diagram does not
contribute to the magnetic moments of the decuplet baryons in the
leading order of the heavy baryon expansion.

For diagram c, there are two adjoint graphs in which the photon
moves from the left vertex to the right one. There are also two
adjoint graphs for diagram f. We include the contributions from the
adjoint graphs in our results. We use diagram f to indicate the
corrections from the renormalization of the external leg where
Lehmann-Symanzik-Zimmermann reduction formula is used.

The leading-order loop contributions to the multipole form
factors are
\begin{eqnarray}
F^{(2,\rm loop)}_{1}&=&\sum_{\phi=\pi,K}\{
\frac{\mathcal{H}^{2}\beta_{\mathcal T}^\phi}{f_\phi^{2}}[
\frac{1}{4}q^{2}(2n^{\rmII0}_{4\phi}+2n^{\rmIII0}_{4\phi})
+\frac{5}{6} n^{\rmIII0}_{13\phi}
+\frac{1}{3}(\frac{q^{2}}{2M_{T}})n^{\rmII0}_{1\phi}
]
+\frac{\mathcal{C}^{2}\beta_{\mathcal N}^\phi}{4f_\phi^{2}}[
2n^{\rmIII}_{13\phi}
+(-\frac{q^{2}}{2M_{T}})n^{\rmII}_{1\phi}]\},\\
F^{(1,\rm loop)}_{2} & = &
\sum_{\phi=\pi,K}\{\frac{\mathcal{H}^{2}\beta_{\mathcal T}^\phi}{f_\phi^{2}}[-\frac{M_{T}}{3}(1+\frac{q^{2}}{4M_{T}^{2}})]n^{\rmII0}_{1\phi}
+\frac{\mathcal{C}^{2}\beta_{\mathcal N}^\phi}{f_\phi^{2}}[-\frac{M_{T}}{2}(1-\frac{q^{2}}{4M_{T}^{2}})]n^{\rmII}_{1\phi}\},\\
F^{(0,\rm loop)}_{3} & = &
\sum_{\phi=\pi,K}\{\frac{\mathcal{H}^{2}\beta_{\mathcal T}^\phi}{f_\phi^{2}}\frac{-1}{3}[
4M_{T}^{2}(2n^{\rmII0}_{4\phi}
+2n^{\rmIII0}_{4\phi})
+4M_{T}n^{\rmII0}_{1\phi}]
+\frac{\mathcal{C}^{2}\beta_{\mathcal N}^\phi}{4f_\phi^{2}}
[4M_{T}^{2}(2n^{\rmII}_{4\phi}+2n^{\rmIII}_{4\phi})+4M_{T}n^{\rmII}_{1\phi}]\},\\
F^{(-1,\rm loop)}_{4} & = &0. \label{eq:F4}
\end{eqnarray}
where $n^{\rmII}_{1\phi}, n^{\rmII}_{4\phi}, n^{\rmIII}_{4\phi},
n^{\rmIII}_{13\phi}$ are $n^{\rmII}_{1}, n^{\rmII}_{4},
n^{\rmIII}_{4}, n^{\rmIII}_{13}$, respectively, defined in the
appendix \ref{appendix-A} with $m=m_\phi$ and $\omega=\delta$. When
$\omega=0$, they become $n^{\rmII0}_{1\phi}, n^{\rmII0}_{4\phi},
n^{\rmIII0}_{4\phi}, n^{\rmIII0}_{13\phi}$. The coefficients
$\beta^\phi_{\mathcal T}$ and $\beta^\phi_{\mathcal N}$ arise from
the decuplet and octet intermediate states respectively. We use the
number $n$ within the parenthesis in the superscript of $X^{(n,
...)}$ to indicate the chiral order of $X$.

The tensor $e \mathcal O_{\rho\mu\sigma}$ at $\mathcal{O}(p^{3})$
should contribute to $F_4$ at $\mathcal{O}(p^{-1})$. However, such
contribution is 0 from Eq. (\ref{eq:F4}). Moreover, all the loop
diagrams in Fig. \ref{fig:allloop} do not contribute to $F_{4}$ up
to $\mathcal{O}(p^{0})$. Therefore, in our case $F_{4}=F_{4}^{(0,
\rm tree)}\sim d_{\mathcal O} Q$. If one tries to obtain the
next-to-leading order correction of $F_4$, $e \mathcal
O_{\rho\mu\sigma}$ at $\mathcal{O}(p^{5})$ must be systematically
considered.

Summing all the contributions in Fig.~\ref{fig:allloop}, the
leading and next-to-leading order loop corrections to
the decuplet magnetic moments can be expressed as
\begin{eqnarray}
\mu_{\mathcal T}^{(2,\rm loop)}& = &
\frac{e}{2M_T}\sum_{\phi=\pi,K}[-\frac{1}{3}\mathcal{H}^{2}M_{T}d_{\mathcal T}\frac{\beta_{\mathcal T}^\phi}{f_\phi^{2}}
-\frac{1}{2}\mathcal{C}^{2}M_{T}\frac{\beta_{\mathcal N}^\phi}{f_\phi^{2}}d_{\mathcal N}],\label{eq:muT2Loop}\\
\mu_{\mathcal T}^{(3,\rm loop)}& = &
\frac{e}{2M_T}\left[\sum_{\phi=\pi,K}(\gamma_{b}^\phi+\gamma_{e}^\phi)\frac{m_\phi^{2}}{8\pi^{2}f_\phi^{2}}\ln\frac{m_\phi}{\mud}
-\sum_{\phi=\pi,K,\eta}(\frac{2}{3f_\phi^{2}}\frac{5b\mathcal{H}^{2}}{12}a_{\mathcal T}\gamma_{a\mathcal T}^\phi +\frac{\mathcal{C}^{2}}{16f_\phi^{2}}a_{\mathcal N}\gamma_{a\mathcal N}^\phi-\frac{b_{2}\mathcal{C}\mathcal{H}}{3\delta f_\phi^{2}}a_{\mathcal T \mathcal N}
 \gamma^\phi_{a\mathcal T \mathcal N})\right]
\nonumber\\&  &
+\sum_{\phi=\pi,K,\eta}(\frac{5\mathcal{H}^{2}}{12}\frac{\mu_{\mathcal T}^{(1)}}{f_\phi^{2}}a_{\mathcal T}\gamma_{f\mathcal T}^\phi
 -\frac{\mu_{\mathcal T}^{(1)}}{4f_\phi^{2}}\mathcal{C}^{2}a_{\mathcal N}\gamma_{f\mathcal N}^\phi),\quad \label{eq:muT3Loop}
\end{eqnarray}
where $\mud=1$ GeV is the renormalization scale. $\gamma^\phi_{a\mathcal T}$, $\gamma^\phi_{a\mathcal N}$, $\gamma^\phi_{a\mathcal T\mathcal N}$,
$\gamma^\phi_{b}$, $\gamma^\phi_{e}$, $\gamma^\phi_{f\mathcal T}$ and
$\gamma^\phi_{f\mathcal N}$ arise from the corresponding diagrams in
Fig.~\ref{fig:allloop}. We collect their explicit expressions in
Tables~\ref{table:beta}, \ref{gHQH2}, \ref{gb} in the Appendix \ref{appendix-C}.

\begin{eqnarray}
d_{\mathcal T} & = & \frac{m_\phi}{16\pi},\\
d_{\mathcal N} & = &\frac{1}{16\pi^2} \begin{cases}\displaystyle
\delta\left(\log\frac{m_\phi^{2}}{\mud^{2}}-1\right)-2\sqrt{\delta^{2}-m_\phi^{2}}\left({\rm arccosh}\left(\frac{-\delta}{m_\phi}\right)-i\pi \right)  & \phi=\pi\\ \displaystyle
\delta\left(\log\frac{m_\phi^{2}}{\mud^{2}}-1\right)+2\sqrt{m_\phi^{2}-\delta^{2}}\arccos\left(-\frac{\delta}{m_\phi}\right)
& \phi=K,\eta
\end{cases},\\
a{}_{\mathcal T} & = & -\frac{m_\phi^{2}}{8\pi^{2}}\ln\frac{m_\phi}{\mud},\\
a{}_{\mathcal N} & = &\frac{1}{16\pi^2} \begin{cases}\displaystyle
\left(m_\phi^{2}-2\delta^{2}\right)\log\left(\frac{m_\phi^{2}}{\mud^{2}}\right)+4\delta\sqrt{\delta^{2}-m_\phi^{2}}
\left({\rm arccosh}\left(\frac{-\delta}{m_\phi}\right)-i\pi\right)+2\delta^{2} & \phi=\pi\\ \displaystyle
\left(m_\phi^{2}-2\delta^{2}\right)\log\left(\frac{m_\phi^{2}}{\mud^{2}}\right)-4\delta\sqrt{m_\phi^{2}-\delta^{2}}\arccos\left(-\frac{\delta}{m_\phi}\right)+2\delta^{2}
& \phi=K,\eta
\end{cases},\\
a_{\mathcal T \mathcal N}&=&\frac{1}{144\pi^{2}}\left[\left(6\delta^{3}-9m_\phi^{2}\delta\right)\log\left(\frac{m_\phi^{2}}{\mud^{2}}\right)+2\left(3\pi m_\phi^{3}+6m_\phi^{2}\delta-5\delta^{3}\right)\right]\nonumber\\
&&-\frac{1}{12\pi^{2}}\begin{cases}\displaystyle
\left(\delta^{2}-m_\phi^{2}\right)^{3/2}\left({\rm arccosh}\left(\frac{-\delta}{m_\phi}\right)-i\pi\right) & \phi=\pi\\ \displaystyle
\left(m_\phi^{2}-\delta^{2}\right)^{3/2}\arccos\left(-\frac{\delta}{m_\phi}\right) & \phi=K,\eta
\end{cases}.
\end{eqnarray}
With the low energy counter terms and loop contributions (\ref{eq:muT2Loop}, \ref{eq:muT3Loop}), we obtain
the magnetic moments,
\begin{equation}
\mu_{\mathcal T}=\left\{\mu_{\mathcal T}^{(1)}\right\}+\left\{\mu_{\mathcal T}^{(2,\rm loop)}\right\}+\left\{\mu_{\mathcal T}^{(3,\rm tree)}+\mu_{\mathcal T}^{(3,\rm loop)}\right\}
\end{equation}
where $\mu_{\mathcal T}^{(1)}$ and $\mu_{\mathcal T}^{(3,\rm tree)}$ is the tree-level magnetic
moments as shown in Table~\ref{Magnetic moments}.

%%%%%%%%%%%%%%%%%%%%%%%%%%%%%%%%%%%%%%%%%%%
\subsection{The electromagnetic form factors and the radii}
%%%%%%%%%%%%%%%%%%%%%%%%%%%%%%%%%%%%%%%%%%%

From the tensor $e \mathcal O_{\rho\mu\sigma}$ up to
$\mathcal{O}(p^{4})$, the magnetic dipole form factor with the
corrections at the next-to-next-leading order is
\begin{equation}
G_{M1}(q^2)=\left\{F_2^{(0)}\right\}
+\left\{F^{(1,\rm loop)}_{2}-F^{(\rm rec, loop)}_{2}\right\}
+\left\{Q\tilde b_{q^2} q^2+F_2^{(2)}+F^{(\rm rec, loop)}_{2}\right\},
\end{equation}
where the terms in the first, second, and the third curly braces are $G_{M1}$ at the leading, next-to-leading, and next-to-next-leading order, respectively. Here
$F_2^{(0)}=2M_T\mu_{\mathcal T}^{(1)}/e$, $F_2^{(2)}=2M_T(\mu_{\mathcal T}^{(3,\rm tree)}+\mu_{\mathcal T}^{(3,\rm loop)})/e$, and
\begin{equation}
F^{(\rm rec, loop)}_{2}=
-\frac{q^{2}}{4M_{T}}\sum_{\phi=\pi,K}(\frac{\mathcal{H}^{2}\beta_{\mathcal T}^\phi}{3f_\phi^{2}}n^{\rmII0}_{1\phi}
-\frac{\mathcal{C}^{2}\beta_{\mathcal N}^\phi}{2f_\phi^{2}}n^{\rmII}_{1\phi}).
\end{equation}

The other multipole form factors are
\begin{eqnarray}
G_{E0}(q^2)&=&\left\{Q\right\}+\left\{Q\tilde c_{r} q^2
+F^{(2,\rm loop)}_{1}-\frac13\tau F^{(0,\rm loop)}_{3} \right\},\\
G_{E2}(q^2)&=&\left\{Q\tilde c_{\mathbb{Q}}-\frac12 F^{(0,\rm loop)}_{3}\right\},\label{eq:GE2}\\
G_{M3}(q^2)&=&Q\tilde d_{\mathcal{O}}.\label{eq:GM3}
\end{eqnarray}
where $\tilde b_{q^2}$, $\tilde c_{r}$, $\tilde c_{\mathbb{Q}}$, and
$\tilde d_{\mathcal{O}}$ are the linear combinations of LECs $b$,
$b_{q^2}$, $c_{r}$, $c_{\mathbb{Q}}$, and $d_{\mathcal{O}}$. We can
estimate the LECs $\tilde b_{q^2}$ and $\tilde c_{r}$ with the SU(3)
VMD model as shown in Section \ref{SecVMD}. However, the LECs
$\tilde c_{\mathbb{Q}}$ and $\tilde d_{\mathcal{O}}$ are still
unknown for the electro quadrupole and magnetic octupole form
factors. Hence we do not list the loop corrections to these
multipole form factors at higher order.

The charge and magnetic radii of the decuplet baryons can be expressed as
\begin{eqnarray}
\langle{r_{E}^{2}}\rangle&=&6\frac{dG_{E0}(q^{2})}{dq^{2}}\mid_{q^{2}=0}
=\langle{r_{E}^{2}}\rangle^{\rm tree}+\langle{r_{E}^{2}}\rangle^{\rm loop}
=[6Q\tilde c_{r}]+ 6[\frac{dF^{(2,\rm loop)}_{1}}{dq^{2}}
\mid_{q^{2}=0}+\frac{1}{12M_{T}^2}F^{(0,\rm loop)}_3(0) ],\\
\langle{r_{M}^{2}}\rangle
&=&\frac{6}{G_{M1}(0)}\frac{dG_{M1}(q^{2})}{dq^{2}} =\langle{r_{M}^{2}}\rangle^{\rm tree}+\langle{r_{M}^{2}}\rangle^{\rm loop}=
\frac{6}{G_{M1}(0)}Q\tilde b_{q^2}+\frac{6}{G_{M1}(0)}\frac{dF^{(1,\rm loop)}_{2}}{dq^{2}}\mid_{q^{2}=0}.
\end{eqnarray}

For the neutral decuplet baryons, we normalize the magnetic radii as
\begin{eqnarray}
\langle{r_{M}^{2}}\rangle
&=&6\frac{dG_{M1}(q^{2})}{dq^{2}}.
\end{eqnarray}

%%%%%%%%%%%%%%%%%%%%%%%%%%%%%%%%%%%%%%%%%%%%%%%%%%%%%%%%
\section{Estimation of the low energy constants} \label{SecEstLEC}
%%%%%%%%%%%%%%%%%%%%%%%%%%%%%%%%%%%%%%%%%%%%%%%%%%%

%%%%%%%%%%%%%%%%%%%%%%%%%%%%%%%%%%%%%%%%%%%%%%%%%%%%%%%%
\subsection{The vector meson dominance model and estimation of some LECs} \label{SecVMD}
%%%%%%%%%%%%%%%%%%%%%%%%%%%%%%%%%%%%%%%%%%%%%%%%%%%

To calculate the tree level charge radii and magnetic radii, we can use the vector meson dominance (VMD) model to estimate the short-distance contribution.

\begin{figure}[tbh]
\centering
\includegraphics[width=0.25\hsize]{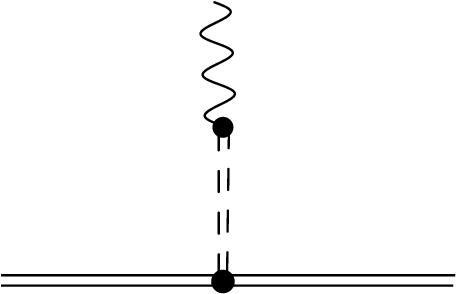}
\caption{The contribution to tree level charge radii from the vector
meson dominance model. The double-solid, the double-dashed and
wiggly lines represent the decuplet baryons, the vector meson and
photon respectively.}\label{fig:rho2}
\end{figure}

It is well-known that the charge radii of the proton and pion are
dominated by the short-distance contribution, which can be estimated
very well by the VMD model. In this work,
we use this model to estimate the LECs $\tilde c_r$ and $\tilde b_{q^2}$ which are related to the
charge and magnetic radii of the decuplet baryons, respectively. Within this framework, the
virtual photon transforms into a virtual vector meson which couples
to the decuplet baryons as shown in Fig.~\ref{fig:rho2}.

It is convenient to adopt the antisymmetric Lorentz tensor field
formulation for the vector meson~\cite{Ecker:1988te,Borasoy:1995ds}, which has six
degrees of freedom. But we can dispose of three of them in a
systematic way. For details see Ref.~\cite{Ecker:1988te}. The
kinetic and mass term of the effective Lagrangian for the vector
meson has the form~\cite{Ecker:1988te,Borasoy:1995ds}
\begin{equation}
\mathcal{L}_{0W}=-\frac{1}{2}{\rm
Tr}(\partial^{\mu}W_{\mu\nu}\partial_{\sigma}W^{\sigma\nu})
+\frac{1}{4}{\rm Tr}(M_{V}^{2}W_{\mu\nu}W^{\mu\nu}),
\end{equation}
where
\begin{equation}
W_{\mu\nu}=\left(\protect\begin{array}{ccc}
\frac{\rho^{0}}{\sqrt{2}}+\frac{\omega}{\sqrt{2}} & \rho^{+} & K^{*+}\protect\\
\rho^{-} & -\frac{\rho^{0}}{\sqrt{2}}+\frac{\omega}{\sqrt{2}} & K^{*0}\protect\\
K^{*-} & \bar{K}^{*0} & \phi \protect\end{array}\right)_{\mu\nu}.
\end{equation}
The QED gauge-invariant interaction between the photon and vector
meson can be written as
\begin{equation}
\mathcal{L}_{W}^{(2)}=\frac{f_{V}}{2\sqrt{2}}{\rm
Tr}(W^{\mu\nu}F_{\mu\nu}^{+}).
\end{equation}
The vector meson and decuplet baryon interaction Lagrangian reads
\begin{equation}
\mathcal{L}_{W\mathcal{T}}^{(1)}=g_{V\mathcal{T}}{\rm Tr}[\bar
T^{\alpha}(\frac{1}{M_{V}}\gamma^{\mu}\nabla^{\nu}W_{\mu\nu}
-\frac{\kappa}{2}\sigma^{\mu\nu}W_{\mu\nu})T_{\alpha}].
\end{equation}
Under the SU(3) symmetry, the charge form factor and charge radii
of the decuplet baryons are
\begin{eqnarray}
G^{\rm VMD}_{E0}(q^{2}) & = & Q\frac{g_{V\mathcal{T}}f_{V}}{\sqrt{2}M_{V}}\frac{q^{2}}{-q^{2}+M_{V}^{2}},\\
\langle r_{E}^{2}\rangle^{\rm tree} & \approx &
\langle r_{E}^{2}\rangle^{\rm VMD}=
6\frac{dG^{\rm VMD}_{E0}(q^{2})}{dq^{2}}|_{q^{2}=0}
=6Q\frac{g_{V\mathcal{T}}f_{V}}{\sqrt{2}}\frac{1}{M_{V}^{3}}.
\end{eqnarray}
The magnetic-dipole form factor and magnetic radii of the decuplet baryons are
\begin{eqnarray}
G^{\rm VMD}_{M1}(q^{2}) & = & Q\frac{g_{V\mathcal{T}}f_{V}}{\sqrt{2}M_{V}}\frac{q^{2}}{-q^{2}+M_{V}^{2}}-\sqrt{2}\kappa Q g_{V\mathcal{T}} f_{V}\frac{M_{T}}{-q^{2}+M_{V}^{2}},\\
\langle r_{M}^{2}\rangle^{\rm tree} & \approx &
\langle r_{M}^{2}\rangle^{\rm VMD}=
\frac{6}{G_{M1}(0)}\frac{dG^{\rm VMD}_{M1}(q^{2})}{dq^{2}}|_{q^{2}=0}
=\frac{6Q}{G_{M1}(0)}[\frac{g_{V\mathcal{T}}f_{V}}{\sqrt{2}M_{V}^{3}}
+\frac{G^{\rm VMD}_{M1}(0)}{M_{V}^{2}Q}].
\end{eqnarray}
Now the LECs $\tilde c_r$ and $\tilde b_{q^2}$ read
\begin{eqnarray}
\tilde c_r&=&\frac{g_{V\mathcal{T}}f_{V}}{\sqrt{2} M_{V}^{3}},\\
\tilde b_{q^2}&=&\frac{g_{V\mathcal{T}}f_{V}}{\sqrt{2}M_{V}^{3}}
+\frac{G^{\rm tree}_{M1}(0)}{M_{V}^{2}Q}.
\end{eqnarray}
In the numerical analysis, we use
$M_{\rho}=770.0\pm0.3$~MeV, $f_{\rho} =f_{V}=152.5\pm16.5$~MeV,
$g_{V\mathcal{T}}\approx
g_{\rho\mathcal{N}}=4.0\pm0.4$~\cite{Kubis:2000zd}, where we have
considered the quark model error around 10\% in Section
\ref{SecQuarkModel}.

%%%%%%%%%%%%%%%%%%%%%%%%%%%%%%%%%%%%%%%%%%%%%%%%%%%%%%%%
\subsection{Quark model and estimation of some couplings} \label{SecQuarkModel}
%%%%%%%%%%%%%%%%%%%%%%%%%%%%%%%%%%%%%%%%%%%%%%%%%%%

Comparing the matrix elements at both the hadron and quark level,
one can express the couplings in terms of the constituent quark
masses and/or other known hadron couplings. To estimate
$g_{V\mathcal{T}}$, we first consider the
$\Delta^{+}\Delta^{+}\rho^{0}$ and $pp\rho^{0}$ vertices at the
hadron level,
\begin{eqnarray}
\mathcal{L}_{\Delta^{+}\Delta^{+}\rho^{0}}&=&\frac{g_{V\mathcal{T}}}{\sqrt{2}M_{V}}{\bar
{\Delta^{+}}}^{\alpha}\gamma^{\mu}\partial^{\nu}\rho^{0}_{\mu\nu}\Delta^{+}_{\alpha},\\
\mathcal{L}_{pp\rho^{0}}&=&\frac{g_{\rho\mathcal{N}}}{\sqrt{2}M_{V}}{\bar p}\gamma^{\mu}\partial^{\nu}\rho^{0}_{\mu\nu}p.
\end{eqnarray}
At the quark level, the quark vector meson interaction reads
\begin{eqnarray}
\mathcal{L}_{qq\rho^{0}}= g_{qq\rho} \bar q_a\gamma^{\mu}\partial^{\nu}\rho^{0}_{\mu\nu}q_a.
\end{eqnarray}
With the help of the flavor wave functions of the static
$\Delta^{+}$ and $p$ states, we obtain the matrix elements at the
hadron level
\begin{eqnarray}
\langle\Delta^{+}\mid i\mathcal{L}_{\Delta^{+}\Delta^{+}\rho^{0}}\mid\Delta^{+};\rho^{0}\rangle &= &
\frac{g_{V\mathcal{T}}}{\sqrt{2}M_{V}}2 m_{\Delta^+} q_{\rho}^\mu \epsilon_{\mu 0} ,\\
\langle p\mid i\mathcal{L}_{pp\rho^{0}}\mid p;\rho^{0}\rangle & = & \frac{g_{\rho\mathcal{N}}}{\sqrt{2}M_{V}}2 m_p q_{\rho}^\mu \epsilon_{\mu 0},
\end{eqnarray}
and at the quark level,
\begin{eqnarray}
\langle\Delta^{+}\mid i\mathcal{L}_{qq\rho^{0}}\mid\Delta^{+};\rho^{0}\rangle
& = &
2g_{qq\rho} (2m_u+m_d) q_{\rho}^\mu \epsilon_{\mu 0},\\
\langle p\mid i\mathcal{L}_{qq\rho^{0}}\mid p;\rho^{0}\rangle
&= &
2g_{qq\rho} (2m_u+m_d) q_{\rho}^\mu \epsilon_{\mu 0},
\end{eqnarray}
Comparing the hadron and quark level matrix
element and neglecting the mass difference between $p$ and $\Delta^+$, we finally obtain
\begin{eqnarray}
g_{V\mathcal{T}}=g_{\rho\mathcal{N}}.
\end{eqnarray}

In the same way, one can estimate the LEC $b_{2}$ by comparing
$\Sigma^{*0}\rightarrow\Lambda+\gamma$ matrix element at both the
hadron and quark level with the Lagrangians
\begin{eqnarray}
\mathcal{L}^{(2)}_{\Sigma^{*0}\rightarrow\Lambda+\gamma}=-\frac{b_{2}}{4M_{B}}{\bar \Sigma^{*0\mu}}
\gamma^{\nu}\gamma_{5}\Lambda F_{\mu\nu},
\end{eqnarray}
and
\begin{eqnarray}
\mathcal{L}_{\rm Im}=-\frac{e}{4}(\frac{2}{3m_{u}}\bar{u}\sigma^{\mu\nu}u-\frac{1}{3m_{d}}\bar{d}\sigma^{\mu\nu}d-\frac{1}{3m_{s}}\bar{s}\sigma^{\mu\nu}s) F_{\mu\nu}.
\end{eqnarray}
We obtain
\begin{eqnarray}
b_{2}=4M_B\sqrt{\frac{3}{2}}(\frac{1}{3\sqrt{6}m_{u}}+\frac{1}{6\sqrt{6}m_{d}})=3.45\pm0.35.
\end{eqnarray}
with $m_{u}=m_{d}=336\pm34~{\rm
MeV}$~\cite{Scadron:2006dy}, where we have considered the quark
model error around 10\%.

%%%%%%%%%%%%%%%%%%%%%%%%%%%%%%%%%%%%%%%%%%%%%%%%%%%%%%%%%%
\section{NUMERICAL RESULTS AND DISCUSSIONS}\label{Sec6}
%%%%%%%%%%%%%%%%%%%%%%%%%%%%%%%%%%%%%%%%%%%%%%%%%%%%%%%%%%

\begin{table}
  \centering
\begin{tabular}{c|c|c|c|c|c}
\toprule[1pt]\toprule[1pt]
 baryons& $\mathcal{O}(p^{1})$ tree & $\mathcal{O}(p^{2})$ loop & $\mathcal{O}(p^{3})$ tree  & $\mathcal{O}(p^{3})$ loop & total \tabularnewline
\midrule[1pt] $\Delta^{++}$ & $\frac{4}{3}b$ & $-3.54$ &
$-\frac{2}{3}d_{1}$ & $0.49-0.50b-0.02b_{D}-0.07b_{F}-0.36g_{2}$ &
4.97(89)\tabularnewline \hline $\Delta^{+}$ & $\frac{2}{3}b$ & $-1.91$ &
$-\frac{2}{3}d_{1}$ & $0.22-0.21b-0.01b_{D}-0.04b_{F}-0.27g_{2}$ &
2.60(50)\tabularnewline \hline $\Delta^{0}$ & 0 & $-0.29$ &
$-\frac{2}{3}d_{1}$ & $-0.27+0.06b+0.001b_{D}-0.001b_{F}-0.18g_{2}$ &
0.02(12)\tabularnewline \hline $\Delta^{-}$ & $-\frac{2}{3}b$ & $1.34$ &
$-\frac{2}{3}d_{1}$ & $-0.32+0.20b+0.01b_{D}+0.02b_{F}-0.14g_{2}$ &
-2.48(32)\tabularnewline \hline $\Sigma^{*+}$ & $\frac{2}{3}b$ & $-1.63$
& $-\frac{2}{3}d_{1}-\frac{2}{9}d_{2}+\frac{4}{9}d_{3}$ &
$0.17-0.50b-0.001b_{D}-0.04b_{F}-0.33g_{2}$ & 1.76(38)\tabularnewline
\hline $\Sigma^{*0}$ & $0$ & $0$ &
$-\frac{2}{3}d_{1}-\frac{2}{9}d_{2}+\frac{1}{9}d_{3}$ &
$-0.02-0.001b_{D}-0.24g_{2}$ & -0.02(3)\tabularnewline \hline
$\Sigma^{*-}$ & $-\frac{2}{3}b$ & $1.63$ &
$-\frac{2}{3}d_{1}-\frac{2}{9}d_{2}-\frac{2}{9}d_{3}$ &
$-0.27+0.50b-0.001b_{D}+0.04b_{F}-0.15g_{2}$ & -1.85(38)\tabularnewline
\hline $\Xi^{*0}$ & $0$ & $0.29$ &
$-\frac{2}{3}d_{1}-\frac{4}{9}d_{2}+\frac{2}{9}d_{3}$ &
$-0.21-0.06b+0.01b_{D}+0.001b_{F}-0.30g_{2}$ & -0.42(13)\tabularnewline
\hline $\Xi^{*-}$ & $-\frac{2}{3}b$ & $1.91$ &
$-\frac{2}{3}d_{1}-\frac{4}{9}d_{2}-\frac{4}{9}d_{3}$ &
$-0.22+0.60b-0.001b_{D}+0.04b_{F}-0.21g_{2}$ &
-1.90(47)\tabularnewline \hline $\Omega^{-}$ & $-\frac{2}{3}b$ & $2.20$
& $-\frac{2}{3}d_{1}-\frac{2}{3}d_{2}-\frac{2}{3}d_{3}$ &
$0.17+0.65b+0.01b_{D}+0.02b_{F}-0.27g_{2}$ & -2.02(5)\tabularnewline
\bottomrule[1pt]\bottomrule[1pt]
\end{tabular}
\caption{The magnetic moments of the decuplet baryons to the next-to-next-leading order (in unit of $\mu_{N}$).} \label{Magnetic moments}
\end{table}

\begin{table}
  \centering
\begin{tabular}{c|c|c|c|c}
\toprule[1pt]\toprule[1pt]
 baryons& $\mathcal{O}(p^{1})$  & $\mathcal{O}(p^{2})$ & $\mathcal{O}(p^{3})$  & PDG\tabularnewline
\midrule[1pt] $\Delta^{++}$ & $4.04(10)$ & $4.90(84)$ & 4.97(89) & 5.6\textpm
1.9\tabularnewline \hline $\Delta^{+}$ & $2.02(5)$ & $2.31(47)$ & 2.60(50) &
2.7 \textpm{} 3.5\tabularnewline \hline $\Delta^{0}$ & 0 & $-0.29(11)$ &
0.02(12)& \tabularnewline \hline $\Delta^{-}$ & $-2.02(5)$ & $-2.88(27)$ &
-2.48(32) & \tabularnewline \hline $\Sigma^{*+}$ & $2.02(5)$ & $2.59(37)$ &
1.76(38) & \tabularnewline \hline $\Sigma^{*0}$ & $0$ & $0$ & -0.02(3)&
\tabularnewline \hline $\Sigma^{*-}$ & $-2.02(5)$ & $-2.59(37)$ & -1.85(38) &
\tabularnewline \hline $\Xi^{*0}$ & $0$ & $0.29(11)$ & -0.42(13) &
\tabularnewline \hline $\Xi^{*-}$ & $-2.02(5)$ & $-2.31(47)$ & -1.90(47) &
\tabularnewline \hline $\Omega^{-}$ & $-2.02(5)$ & $-2.02(5)$ & -2.02(5) &
-2.02 \textpm 0.05\tabularnewline \bottomrule[1pt]\bottomrule[1pt]
\end{tabular}
\caption{The magnetic moments of the decuplet baryons when the
chiral expansion is truncated at $\mathcal{O}(p^{1})$,
$\mathcal{O}(p^{2})$ and $\mathcal{O}(p^{3})$ respectively (in unit
of $\mu_{N}$). } \label{various orders Magnetic moments}
\end{table}

We collect our numerical results of the magnetic moments of the
decuplet baryons to the next-to-next-leading order in Table~\ref{Magnetic
moments}. We also compare the numerical results of the magnetic
moments when the chiral expansion is truncated at orders
$\mathcal{O}(p^{1})$, $\mathcal{O}(p^{2})$ and $\mathcal{O}(p^{3})$
respectively in Table~\ref{various orders Magnetic moments}.

At the leading order $\mathcal{O}(p^{1})$, there is only one unknown
low energy constant $b$. We use the precise experimental measurement
of the $\Omega^{-}$ magnetic moment
$\mu_{\Omega^{-}}=(-2.02\pm0.05)\mu_{N}$ as input to extract
$b=3.03\pm0.08$. The magnetic moments of the other decuplet baryons are
given in the second column in Table~\ref{various orders Magnetic
moments}. Notice that the $\mathcal{O}(p^{1})$ tree level magnetic
moments of the neutral baryons $\Delta^{0}$, $\Sigma^{*0}$ and
$\Xi^{*0}$ vanish. In the limit of the exact SU(3) flavor symmetry,
there exits only one independent term for the magnetic interaction
in the $\mathcal{O}(p^{2})$ Lagrangian of the decuplet baryons due
to the constraint of the decuplet flavor structure. Therefore, the
leading order $\mathcal{O}(p^{1})$ magnetic moments of the decuplet
baryons are proportional to their charge, which is in strong
contrast with the case of the octet baryons. The magnetic moments of
the neutral octet baryons do not vanish at the leading order because
there exist two independent magnetic interaction terms as
illustrated in Refs.~\cite{Jenkins:1992pi,Meissner:1997hn}.

Up to $\mathcal{O}(p^{2})$, we need include both the leading
tree-level magnetic moments and the $\mathcal{O}(p^{2})$ loop
corrections. At this order, all the coupling constants are
well-known. There do not exist new LECs. Again, we use the
experimental value of the $\Omega^{-}$ magnetic moment
$\mu_{\Omega^{-}}=(-2.02\pm0.05)\mu_{N}$ as input to extract the LEC
$b=6.3\pm0.1$. We list the numerical results in the
third column in Table~\ref{various orders Magnetic moments}, where
the errors in the brackets are dominated by the errors of the
coupling constants $\mathcal{C}, \mathcal{H}$ in Eq.
(\ref{Eq:baryon2}).

It's interesting to notice that the magnetic moment of $\Sigma^{*0}$
still vanishes even at $\mathcal{O}(p^{2})$. The reason is as follows.
Throughout our calculation, we neglect the mass difference among
different decuplet baryons in the loop and have used the same
propagator $\frac{-iP_{\rho\sigma}^{3/2}}{v\cdot q+i\epsilon}$ for all the
decuplet baryons. In the case of the $\Sigma^{*0}$ magnetic moment,
the loop contributions from different intermediate states cancel
each other. I.e., the pion loop contributions with the intermediate
baryons $\Sigma^{*+}$ and $\Sigma^{*-}$, $\Sigma^{+}$ and
$\Sigma^{-}$ cancel each other due to the exact SU(2) flavor
symmetry. The kaon loop contributions with the intermediate baryons
$\Delta^{+}$ and $\Xi^{*-}$, $p$ and $\Xi^{-}$ cancel each other due
to the SU(3) flavor symmetry. Hence, the magnetic moment of
$\Sigma^{*0}$ is zero to $\mathcal{O}(p^{2})$ in Table~\ref{various orders Magnetic moments}.

Up to $\mathcal{O}(p^{3})$, there are seven unknown LECs: $b_{D,F}$,
$b$, $g_{2}$, $d_{1,2,3}$. The first two LECs were extracted in the
calculation of the magnetic moments of the octet baryons in
Ref.~\cite{Meissner:1997hn}: $b_{D}=3.9$, $b_{F}=3.0$. We use the experimental value of the $\Omega^{-}$
magnetic moment, the magnetic moments of the $\Delta$ baryons in
Ref.~\cite{Cloet:2003jm} ($\mu _{\Delta^{++}}=4.99\pm0.56,\mu
_{\Delta^{+}}=2.49\pm0.27,\mu _{\Delta^{0}}=0.06\pm0.00,\mu
_{\Delta^{-}}=-2.45\pm0.27$) and $\mu _{\Sigma^{*0}}=0$ to extract
the remaining five LECs: $b=6.8\pm0.4$, $g_{2}=-13.7\pm0.1$,
$d_{1}=3.5\pm0.1$, $d_{2}=-1.5\pm0.1$, $d_{3}=4.3\pm0.1$. We list
the numerical results up to $\mathcal{O}(p^{3})$ in the fourth
column in Table~\ref{various orders Magnetic moments} after taking
the uncertainties of these inputs into consideration. In the error
analysis, we use the least $\chi^2$ fitting tool of the TMinuit
software package to get the errors of fitting. To get the total
errors of the $\mathcal{O}(p^{3})$ magnetic moments, we have
considered the errors of the coupling constants $\mathcal{C},
\mathcal{H}$, the error of coupling constant $b_2$ and the errors of
fitting.

In order to study the convergence of the chiral expansion, we show
the numerical results at each order for the decuplet magnetic
moment:
\begin{eqnarray}
\mu_{\Delta^{++}}&=&9.0(1-0.39-0.06)=4.97,\nonumber\\
\mu_{\Delta^{+}}&=&4.5(1-0.43+0.01)=2.60,\nonumber\\
\mu_{\Delta^{0}}&=&-0.29(0+1-1.06)=0.02,\nonumber\\
\mu_{\Delta^{-}}&=&-4.5(1-0.30-0.15)=-2.48,\nonumber\\
\mu_{\Sigma^{*+}}&=&4.5(1-0.36-0.25)=1.76,\nonumber\\
\mu_{\Sigma^{*0}}&=&0+0-0.02,\nonumber\\
\mu_{\Sigma^{*-}}&=&-4.5(1-0.36-0.23)=-1.85,\nonumber\\
\mu_{\Xi^{*0}}&=&0.29(0+1-2.44)=-0.42,\nonumber\\
\mu_{\Xi^{*-}}&=&-4.5(1-0.43-0.15)=-1.90,\nonumber\\
\mu_{\Omega^{-}}&=&-4.5(1-0.49-0.06)=-2.02.
\end{eqnarray}
For the neutral decuplet baryons, their magnetic moments vanish at
$\mathcal{O}(p^{1})$. Their total magnetic moments arise from the
loop contributions at $\mathcal{O}(p^{2,3})$ and the tree-level LECs
$d_{1,2,3}$ at $\mathcal{O}(p^{3})$ which are related to the strange
quark mass correction. For the charged baryons, one observes rather
good convergence of the chiral expansion and the leading order term
dominates in these channels.

\begin{figure}
\centering
\includegraphics[width=0.6\hsize]{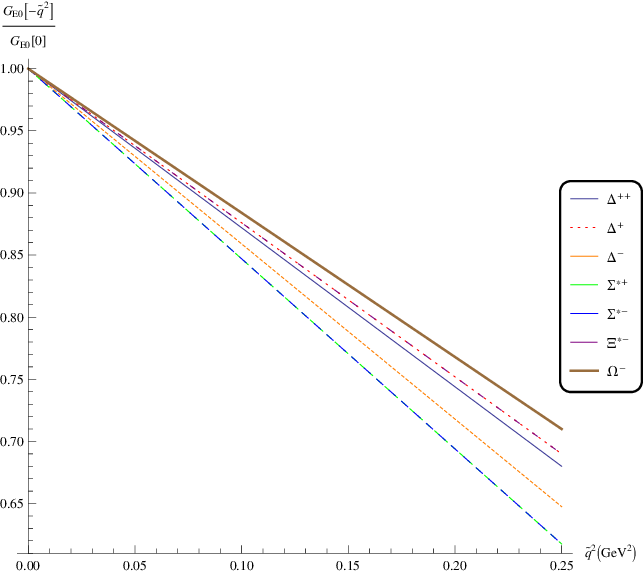}
\label{fig:side:a} \caption{The variation of the normalized electric
charge form factor $G_{E0}(-\tilde q^2)$ with $\tilde q^2=-q^2>0$.}
\label{fig:GE0_abso}
\end{figure}

\begin{figure}[tbh]
\centering
\includegraphics[width=0.6\hsize]{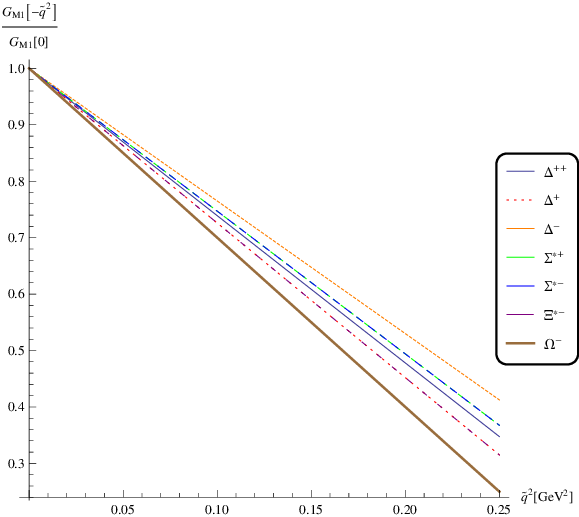}
\caption{ The variations of ${G_{M1}(-\tilde q^2)\over
G_{M1}(0)}$ with $\tilde q^2$. }\label{fig:GM1_proportion}
\end{figure}

In order to illustrate the variation of the multipole form factors
with the photon momentum $q$, we show the $\tilde q^2=-q^2$
dependence of the electric charge and magnetic dipole form factors
to $\mathcal{O}(p^{3})$ in
Figs.~\ref{fig:GE0_abso}-\ref{fig:GM1_proportion}, where we have
used the SU(3) VMD model to estimate the LEC $b_{q^2}$ and $c_{r}$
as shown in Eqs.~(\ref{Eq:baryon3}),(\ref{Eq:chargeradii}).

In Fig. \ref{fig:GE0_abso} or Fig. \ref{fig:GM1_proportion}, we
notice that there is not much difference between the slopes of the
curves. They should be exactly the same for different decuplet
baryons if only the tree-level contributions are considered. The
difference arises from the loop correction.

\begin{figure}[tbh]
\centering
\includegraphics[width=0.6\hsize]{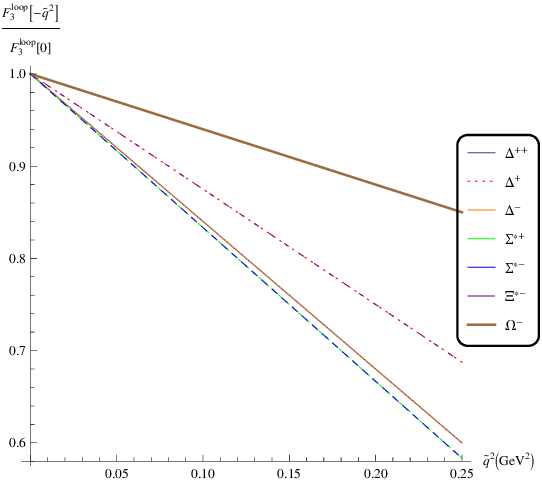}
\caption{The variations of ${F^{(0,\rm loop)}_{3}(-\tilde q^2)\over
F^{(0,\rm loop)}_{3}(0)}$ with $\tilde q^2$.}\label{fig:GE2proportion}
\end{figure}

The electric quadrupole form factors $G_{E2}(q^2)$ contain
interesting information on the deformation of decuplet baryons.
$\tilde c_\mathbb{Q}$ cannot be determined because of the lack of
experimental data. But the $\tilde c_\mathbb{Q}$ term does not
change with $q^2$. We list the normalized $F^{(0,\rm
loop)}_{3}(-\tilde q^2)$ in Fig. \ref{fig:GE2proportion} to indicate
the variation of $G_{E2}(q^2)$.

In Table~\ref{table:radii} we show numerical results for the charge
radii and magnetic radii of the decuplet baryons. One can check that
the charge radii estimated from the VMD model are proportional to
the charge $Q$ of the decuplet baryons, while the magnetic radii
estimated from the VMD model are the same for different baryons.
In the error analysis, the errors of VMD radii are
dominated by the input parameters $M_{\rho}, f_{V},
g_{V\mathcal{T}}$ and their propagation. The chiral correction radii
are dominated by the errors of the coupling constants $\mathcal{C},
\mathcal{H}$ in Eq. (\ref{Eq:baryon2}).

\begin{table}
\centering
\begin{tabular}{c|c|c|c||c|c|c|c}
\toprule[1pt]\toprule[1pt]
$\langle r_{E}^{2}\rangle/\rm fm^{2}$ & VMD & chiral correction & total value&$\langle r_{M}^{2}\rangle/\rm fm^{2}$ & VMD & chiral correction & total value\tabularnewline
\hline
$\Delta^{++}$ & 0.44(20) &0.16(6) & 0.60(21)&$\Delta^{++}$ & 0.46(11) & 0.15(10) & 0.61(15)\tabularnewline
\hline
$\Delta^{+}$ & 0.22(10) & 0.07(3)&0.29(10)& $\Delta^{+}$ & 0.46(11) & 0.18(8) & 0.64(14)\tabularnewline
\hline
$\Delta^{0}$ & 0 &-0.02(1)  & -0.02(1)&$\Delta^{0}$ & 0 & 0.07(12) & 0.07(12)  \tabularnewline
\hline
$\Delta^{-}$ & -0.22(10) &-0.11(5) &-0.33(11)& $\Delta^{-}$ & 0.46(11) & 0.09(15) &0.55(19) \tabularnewline
\hline
$\Sigma^{*+}$ &0.22(10)  &0.09(4)  & 0.31(11)& $\Sigma^{*+}$ & 0.46(11) & 0.13(12) & 0.59(16)\tabularnewline
\hline
$\Sigma^{*0}$ & 0 &0 & 0& $\Sigma^{*0}$ & 0 & 0 & 0\tabularnewline
\hline
$\Sigma^{*-}$ &-0.22(10) & -0.09(4) &-0.31(11)& $\Sigma^{*-}$ & 0.46(11) & 0.13(12) & 0.59(16) \tabularnewline
\hline
$\Xi^{*0}$ & 0 & 0.02(1) & 0.02(1)&$\Xi^{*0}$ & 0 & -0.07(12) & -0.07(12) \tabularnewline
\hline
$\Xi^{*-}$ & -0.22(10) & -0.07(3) &-0.29(10)&  $\Xi^{*-}$ & 0.46(11) & 0.18(8) & 0.64(14)\tabularnewline
\hline
$\Omega^{-}$ & -0.22(10) &-0.05(2) & -0.27(10)& $\Omega^{-}$ & 0.46(11) & 0.24(4) & 0.70(12)\tabularnewline
\bottomrule[1pt]\bottomrule[1pt]
\end{tabular}
 \caption{Charge radii and magnetic radii (in ${\rm fm^{2}}$).}
\label{table:radii}
\end{table}

%%%%%%%%%%%%%%%%%%%%%%%%%%%%%%%%%%%%%%%%%%%%%%%%%%%%%%%%%%
\section{Conclusions}\label{Sec7}
%%%%%%%%%%%%%%%%%%%%%%%%%%%%%%%%%%%%%%%%%%%%%%%%%%%%%%%%%%

In short summary, we have systematically studied the magnetic
moments of the decuplet baryons up to the next-to-next-leading order in the
framework of the heavy baryon chiral perturbation theory. With both
the octet and decuplet baryon intermediate states in the chiral
loops, we have systematically calculated the chiral corrections to
the magnetic moments of the decuplet baryons order by order. The
chiral expansion converges rather well for the charged channels. In Table~\ref{Comparison of magnetic moments}, we compare our results obtained in the HBChPT with those from other model
calculations such as lattice QCD~\cite{Leinweber:1992hy}, chiral quark model~\cite{Wagner:2000ii}, non relativistic quark model~\cite{Hikasa:1992je}, QCD sum rules
~\cite{Lee:1997jk}, large $N_{c}$~\cite{Luty:1994ub}, covariant ChPT~\cite{Geng:2009ys} and next-to-leading order HBChPT~\cite{Butler:1993ej}. We also list the
experimental values in the PDG~\cite{Agashe:2014kda}. One may
observe the qualitatively similar features for the magnetic moments
of the decuplet baryons.

Because of the SU(3) flavor symmetry, there is one independent low
energy constant at the leading order. Hence, the magnetic moments of
the decuplet baryons are proportional to their charge. Therefore,
the magnetic moments of the neutral decuplet baryons vanish at
$\mathcal{O}(p^{1})$, which differs from the case of the neutral
octet baryons. There exist two independent magnetic interaction
terms for the octet baryons, which ensures a large magnetic moment
for the neutron at the leading order.

For the magnetic moment of the $\Sigma^{*0}$, the pion loop
contributions with the $\Sigma^{*+}$ and $\Sigma^{*-}$, $\Sigma^{+}$
and $\Sigma^{-}$ intermediate states cancel each other exactly in
the SU(2) symmetry limit. The kaon loop contributions with the
$\Delta^{+}$ and $\Xi^{*-}$, $p$ and $\Xi^{-}$ intermediate states
cancel each other exactly in the SU(3) symmetry limit. The magnetic
moment of $\Sigma^{*0}$ vanishes even at $\mathcal{O}(p^{2})$ with
SU(3) symmetry. The non-vanishing SU(3) breaking corrections first
appear at $\mathcal{O}(p^{3})$. In other words, the SU(3) flavor
symmetry demands that the magnetic moment of $\Sigma^{*0}$ be
significantly smaller than those of the charged decuplet baryons.

We hope that the magnetic moments of the decuplet baryons will be
measured experimentally in future experiments. Moreover, the
analytical expressions derived in this work may be useful to the
possible chiral extrapolation of the lattice simulations of the
decuplet electromagnetic properties in the coming future.

 \begin{table}
  \centering
\begin{tabular}{c|c|c|c|c|c|c|c|c|c|c}
\toprule[1pt]\toprule[1pt] baryons & $\Delta^{++}$ & $\Delta^{+}$ &
$\Delta^{0}$ & $\Delta^{-}$ & $\Sigma^{*+}$ & $\Sigma^{*0}$ &
$\Sigma^{*-}$ & $\Xi^{*0}$ & $\Xi^{*-}$ &
$\Omega^{-}$\tabularnewline \midrule[1pt]
LQCD~\cite{Leinweber:1992hy} & 6.09 & 3.05 & 0  & -3.05 & 3.16 &
0.329& -2.50 & 0.58 & -2.08 & -1.73\tabularnewline \hline
ChQM~\cite{Wagner:2000ii} & 6.93 & 3.47 & 0 & -3.47 & 4.12 & 0.53 &
-3.06 & 1.10 & -2.61 & -2.13\tabularnewline \hline
NQM~\cite{Hikasa:1992je} & 5.56 & 2.73 & -0.09 & -2.92 & 3.09 &
0.27 & -2.56 & 0.63  & -2.2 &  -1.84\tabularnewline \hline
QCD-SR~\cite{Lee:1997jk}& 4.1& 2.07 & 0  & -2.07 & 2.13& -0.32& -1.66
& -0.69& -1.51 & -1.49\tabularnewline \hline
large $N_{c}$~\cite{Luty:1994ub}
& 5.9& 2.9 & \textemdash{}  & -2.9& 3.3& 0.3& -2.8& 0.65 & -2.30 &
-1.94\tabularnewline \hline
covariant ChPT~\cite{Geng:2009ys} & 6.04& 2.84 &
-0.36& -3.56& 3.07& 0  & -3.07 & 0.36 & -2.56&  -2.02\tabularnewline
\hline
HBChPT~\cite{Butler:1993ej}& 4.0 & 2.1& -0.17& -2.25 & 2.0&
-0.07 & -2.2 & 0.10 & -2.0 & -1.94\tabularnewline \hline
PDG~\cite{Agashe:2014kda}  & 5.6\textpm 1.9 & 2.7 $\pm$ 3.5 &
\textemdash{}  & \textemdash{}  & \textemdash{}  & \textemdash{}  &
\textemdash{}  & \textemdash{}  & \textemdash{}  & -2.02
$\pm$0.05\tabularnewline \midrule[1pt] this work & 4.97(89)& 2.60(50) & 0.02(12)
& -2.48(32)& 1.76(38)& -0.02(3) & -1.85(38) & -0.42(13) & -1.90(47) & -2.02(5)\tabularnewline
\bottomrule[1pt]\bottomrule[1pt]
\end{tabular}
\caption{Comparison of the magnetic moments of the decuplet baryons
in literature including lattice QCD(LQCD)~\cite{Leinweber:1992hy}, chiral quark model(ChQM)~\cite{Wagner:2000ii}, non relativistic quark model(NQM)~\cite{Hikasa:1992je}, QCD sum rules(QCD-SR)~\cite{Lee:1997jk}, large $N_{c}$~\cite{Luty:1994ub}, covariant ChPT~\cite{Geng:2009ys}, next-to-leading order HBChPT~\cite{Butler:1993ej} and PDG~\cite{Agashe:2014kda}(in unit of $\mu_{N}$).}
  \label{Comparison of magnetic moments}
 \end{table}

\section*{ACKNOWLEDGMENTS}

H. S. Li is very grateful to N. Jiang, B. Zhou, L. Ma, and G. J. Wang
for very helpful discussions. This project is supported by the
National Natural Science Foundation of China under Grants
11575008, 11621131001 and 973 program.

\begin{appendix}

%%%%%%%%%%%%%%%%%%%%%%%%%%%%%%%%%%%%%%%%%%%%%%%%%%%%%%
\section{Integrals and loop functions} \label{appendix-A}
%%%%%%%%%%%%%%%%%%%%%%%%%%%%%%%%%%%%

We collect some common integrals and loop functions in this
appendix.

\subsection{The integrals with one or two meson propagators}
\begin{eqnarray}
\Delta & = & i\int\frac{d^{d}l \,\mud^{4-d}}{(2\pi)^{d}}\frac{1}{l^{2}-m^{2}}= 2m^{2}(L(\mud)+\frac{1}{32\pi^{2}}\ln\frac{m^{2}}{\mud^{2}}),\\
L(\mud) & = &
\frac{\mud^{d-4}}{16\pi^{2}}[\frac{1}{d-4}-\frac{1}{2}(\ln4\pi+1+\Gamma^{\prime}(1))].
\end{eqnarray}
\begin{eqnarray}
I_{0}(q^{2})&=&i\int\frac{d^{d}l\,\mud^{4-d}}{(2\pi)^{d}}\frac{1}{(l^{2}-m^{2}+i\epsilon)((l+q)^{2}-m^{2}+i\epsilon)}\nonumber\\
&=&\begin{cases} \displaystyle
-\frac{1}{16\pi^{2}}(1-\ln\frac{m^{2}}{\mud^{2}}-r\ln|\frac{1+r}{1-r}|)+2L(\mud) & \left(q^{2}<0\right)\\ \displaystyle
-\frac{1}{16\pi^{2}}(1-\ln\frac{m^{2}}{\mud^{2}}-2r\, \arctan\frac{1}{r})+2L(\mud) & (0<q^{2}<4m^{2})\\ \displaystyle
-\frac{1}{16\pi^{2}}(1-\ln\frac{m^{2}}{\mud^{2}}-r\ln|\frac{1+r}{1-r}|+i\pi
r)+2L(\mud) & (q^{2}>4m^{2})
\end{cases},
\end{eqnarray}
where $r=\sqrt{|1-4m^{2}/q^{2}|}$.

\subsection{The integrals with one baryon propagator and one meson
propagator}

\begin{equation}
i\int\frac{d^{d}l \,\mud^{4-d}}{(2\pi)^{d}}\frac{[1,l_{\alpha},l_{\alpha}l_{\beta}]}{(l^{2}-m^{2}+i\epsilon)(\omega+v\cdot
l+i\epsilon)}
=[J_{0}(\omega),v_{\alpha}J_{1}(\omega),g_{\alpha\beta}J_{2}(\omega)+v_{\alpha}v_{\beta}J_3(\omega)],
\omega=v\cdot r+\delta
\end{equation}
\begin{equation}
J_{0}(\omega)=\begin{cases}
\displaystyle
\frac{-\omega}{8\pi^{2}}(1-\ln\frac{m^{2}}{\mud^{2}})+\frac{\sqrt{\omega^{2}-m^{2}}}{4\pi^{2}}({\rm arccosh}\frac{\omega}{m}-i\pi)+4\omega L(\mud) & (\omega>m)\\ \displaystyle
\frac{-\omega}{8\pi^{2}}(1-\ln\frac{m^{2}}{\mud^{2}})+\frac{\sqrt{m^{2}-\omega^{2}}}{4\pi^{2}}\arccos\frac{-\omega}{m}+4\omega L(\mud) & (\omega^{2}<m^{2})\\ \displaystyle
\frac{-\omega}{8\pi^{2}}(1-\ln\frac{m^{2}}{\mud^{2}})-\frac{\sqrt{\omega^{2}-m^{2}}}{4\pi^{2}}{\rm
arccosh}\frac{-\omega}{m}+4\omega L(\mud) & (\omega<-m)
\end{cases}
\end{equation}

\begin{eqnarray}
J_{1}(\omega)&=&-\omega J_{0}(\omega)+\Delta\\
J_2(\omega)&=&\frac{1}{d-1}[(m^2-\omega^2)J_0(\omega)+\omega\Delta]\\
J_3(\omega)&=&-\omega J_1(\omega)-J_2(\omega)
\end{eqnarray}

\subsection{The integrals with two baryon propagators and one meson propagator}

\begin{eqnarray}
i\int\frac{d^{d}l \,\mud^{4-d}}{(2\pi)^{d}}\frac{[1,l_{\alpha},l_{\alpha}l_{\beta}]}{(l^{2}-m^{2}+i\epsilon)(v\cdot
l+i\epsilon)(\omega+v\cdot l+i\epsilon)}
=[\Gamma_{0}(\omega),v_{\alpha}\Gamma_{1}(\omega),g_{\alpha\beta}\Gamma_{2}(\omega)+v_{\alpha}v_{\beta}\Gamma_3(\omega)]
\qquad \omega\neq 0
\end{eqnarray}

\begin{eqnarray}
\Gamma_{i}(\omega)=\frac{1}{\omega}[J_i(0)-J_i(\omega)]
\end{eqnarray}

\begin{eqnarray}
i\int\frac{d^{d}l\,\mud^{4-d}}{(2\pi)^{d}}\frac{[1,l_{\alpha},l_{\alpha}l_{\beta}]}{(l^{2}-m^{2}+i\epsilon)(\omega+v\cdot
l+i\epsilon)^2} =-[\frac{\partial}{\partial
\omega}J_{0}(\omega),v_{\alpha}\frac{\partial}{\partial
\omega}J_{1}(\omega),g_{\alpha\beta}\frac{\partial}{\partial
\omega}J_{2}(\omega)+v_{\alpha}v_{\beta}\frac{\partial}{\partial
\omega}J_3(\omega)]
\end{eqnarray}

\subsection{The integrals with one baryon propagator and two meson propagators}

\begin{eqnarray*}
i\int\frac{d^{d}l\,\mud^{4-d}}{(2\pi)^{d}}\frac{[1,l_{\alpha},l_{\alpha}l_{\beta},l_{\nu}l_{\alpha}l_{\beta}]}{(l^{2}-m^{2}+i\epsilon)((l+q)^{2}-m^{2}+i\epsilon)(\omega+v\cdot
l+i\epsilon)} & = &
[L_{0}(\omega),L_{\alpha},L_{\alpha\beta},L_{\nu\alpha\beta}],v\cdot q=0
\end{eqnarray*}

\begin{eqnarray}
L_{0}(\omega)& = & \protect\begin{cases} \displaystyle
\frac{-1}{8\pi^{2}}\frac{1}{\sqrt{\omega^{2}-m^{2}}}({\rm arccosh}\frac{\omega}{m}-i\pi) & (\omega>m)\\ \displaystyle
\frac{1}{8\pi^{2}}\frac{1}{\sqrt{m^{2}-\omega^{2}}}\arccos\frac{-\omega}{m} & (\omega^{2}<m^{2})\protect\\ \displaystyle
\frac{1}{8\pi^{2}}\frac{1}{\sqrt{\omega^{2}-m^{2}}}{\rm
arccosh}\frac{-\omega}{m} & (\omega<-m)
\protect\end{cases}.\\
L_{\alpha}& = & n^{\rmI}_{1}q_{\alpha}+n^{\rmI}_{2}v_{\alpha}\\
L_{\alpha\beta} & = &
n^{\rmII}_{1}g_{\alpha\beta}+n^{\rmII}_{2}q_{\alpha}q_{\beta}+n^{\rmII}_{3}v_{\alpha}v_{\beta}
 +n^{\rmII}_{4}v_{\alpha}q_{\beta}+n^{\rmII}_{5}q_{\alpha}v_{\beta}\\
 L_{\nu\alpha\beta} & = & n^{\rmIII}_{1}q_{\nu}q_{\alpha}q_{\beta}+n^{\rmIII}_{2}q_{\nu}q_{\alpha}v_{\beta}
 +n^{\rmIII}_{3}q_{\nu}q_{\beta}v_{\alpha}+n^{\rmIII}_{4}q_{\alpha}q_{\beta}v_{\nu}
 +n^{\rmIII}_{5}q_{\nu}g_{\alpha\beta}\nonumber\\
 &  & +n^{\rmIII}_{6}q_{\beta}g_{\nu\alpha}+n^{\rmIII}_{7}q_{\alpha}g_{\nu\beta}+n^{\rmIII}_{8}q_{\nu}v_{\alpha}v_{\beta}
 +n^{\rmIII}_{9}q_{\alpha}v_{\nu}v_{\beta}
 +n^{\rmIII}_{10}q_{\beta}v_{\nu}v_{\alpha}\nonumber\\
 &  & +n^{\rmIII}_{11}g_{\nu\beta}v_{\alpha}+n^{\rmIII}_{12}g_{\nu\alpha}v_{\beta}
 +n^{\rmIII}_{13}g_{\alpha\beta}v_{\nu}+n^{\rmIII}_{14}v_{\nu}v_{\alpha}v_{\beta}
 \end{eqnarray}

\subsection{The explicit expressions of the scalar functions}

\begin{eqnarray*}
n^{\rmI}_{1}& = & -\frac{L_{0}}{2}\\
n^{\rmI}_{2}& = & I_{0}-L_{0} \omega\\
n^{\rmII}_{1}& = &\frac{-4 I_{0} \omega -2 J_{0}+q^2 L_{0}-4 L_{0} m^2+4 L_{0} \omega ^2}{8-4 d}\\
n^{\rmII}_{2}& = &\frac{2 (d-3) J_{0}+(d-1) q^2 L_{0}-4 \left(I_{0} \omega +L_{0} m^2-L_{0} \omega ^2\right)}{4 (d-2) q^2}\\
n^{\rmII}_{3}& = &\frac{-4 \left(d I_{0}\omega -d L_{0} \omega ^2-I_{0} \omega +L_{0} m^2+L \omega ^2\right)-2 J_{0}+q^2 L_{0}}{4 (d-2)}\\
n^{\rmII}_{4}& = &n^{\rmII}_{5}=\frac{1}{2} (L_{0} \omega -I_{0})\\
n^{\rmIII}_{1}& = &-\frac{1}{8 (d-2) q^2}[6 (d-3) J_{0}+(d+1) q^2 L_{0}-12 \left(I_{0} \omega +L_{0} m^2-L_{0} \omega ^2\right)]\\
n^{\rmIII}_{2,3,4}& = &\frac{1}{4 (d-2) (d-1) q^2}[d^2 q^2 \left(I_{0} - L_{0} \omega \right)-2 \left(d^2-4 d+3\right)\omega J_{0} \\
&&-2 d \left(\Delta +I_{0} \left(q^2-2 \omega ^2\right)+L_{0} \omega  \left(-2 m^2+2 \omega ^2-q^2\right)\right)\\
&&+4 \Delta +4 I_{0} m^2-4 I_{0} \omega ^2-q^2 L_{0} \omega -4 L_{0} m^2 \omega +4 L_{0} \omega ^3]\\
n^{\rmIII}_{5,6,7}&=&\frac{1}{8 (d-2)}[-4 I_{0} \omega -2 J_{0}+q^2 L_{0}-4 L_{0} m^2+4 L_{0} \omega ^2]\\
n^{\rmIII}_{8,9,10}&=&\frac{1}{16-8 d}[-4 d I_{0} \omega +4 d L_{0} \omega ^2+4 I_{0} \omega -2 J_{0}+q^2 L_{0}-4 L_{0} m^2-4 L_{0} \omega ^2]\\
n^{\rmIII}_{11,12,13}&=&\frac{1}{4 (d-2) (d-1)}[4 d \Delta +I_{0} \left(4 \left((d-3) m^2-(d-1) \omega ^2\right)-(d-2)q^2\right)\\
&&-2 (d-1) \omega  J_{0}+d q^2 L_{0} \omega -4 d L_{0} m^2 \omega +4 d L_{0} \omega ^3-8 \Delta -q^2 L_{0} \omega+4 L_{0} m^2 \omega -4 L_{0} \omega ^3]\\
n^{\rmIII}_{14}&=&\frac{1}{4 (d-2) (d-1)}[2 I_{0} \left(2 \left(d^2-1\right) \omega ^2+(d-2) q^2+2 (7-2d) m^2\right)-4 d^2 L_{0} \omega ^3\\
&&-10 d \Delta +6 (d-1) \omega  J_{0}-3 d q^2 L_{0} \omega +12 d L_{0} m^2 \omega +20 \Delta +3 q^2 L_{0} \omega -12 L_{0} m^2 \omega +4 L_{0} \omega ^3]
 \end{eqnarray*}

%%%%%%%%%%%%%%%%%%%%%%%%%%%%%%%%%%%%%%%%%%%%%%%%%%%%%%%%
\section{THE COEFFICIENTS OF THE LOOP CORRECTIONS} \label{appendix-C}

In this appendix, we collect the explicit formulae for the chiral expansion of the decuplet baryon
magnetic moments at $\mathcal{O}(p^{2})$ in Table \ref{table:beta}  and $\mathcal{O}(p^{3})$ in Tables \ref{gHQH2} and \ref{gb}, respectively.

\begin{table}
  \centering
\begin{tabular}{c|c|c|c|c|c|c}
\toprule[1pt]\toprule[1pt] baryons & $\beta_{\mathcal T}^{\pi}$ &
$\beta_{\mathcal T}^{K}$ & $\beta_{\mathcal T}^{\eta}$ & $\beta_{\mathcal N}^{\pi}$ &
$\beta_{\mathcal N}^{K}$ & $\beta_{\mathcal N}^{\eta}$\tabularnewline
\midrule[1pt] $\Delta^{++}$ & $\frac{2}{3}$ &
$\frac{2}{3}$ & 0 & $2$ &
$2$ & 0\tabularnewline \hline $\Delta^{+}$ &
$\frac{2}{9}$ & $\frac{4}{9}$ & 0 &
$\frac{2}{3}$ & $\frac{4}{3}$ &
0\tabularnewline \hline $\Delta^{0}$ & $-\frac{2}{9}$
& $\frac{2}{9}$ & 0 & $-\frac{2}{3}$ &
$\frac{2}{3}$ & 0\tabularnewline \hline $\Delta^{-}$
& $-\frac{2}{3}$ & 0 & 0 & $-2$ & 0 &
0\tabularnewline \hline $\Sigma^{*+}$ & $\frac{4}{9}$
& $\frac{2}{9}$ & 0 & $\frac{4}{3}$ &
$\frac{2}{3}$ & 0\tabularnewline \hline $\Sigma^{*0}$
& 0 & 0 & 0 & 0 & 0 & 0\tabularnewline \hline $\Sigma^{*-}$ &
$-\frac{4}{9}$ & $-\frac{2}{9}$ & 0 &
$-\frac{4}{3}$ & $-\frac{2}{3}$ &
0\tabularnewline \hline $\Xi^{*0}$ & $\frac{2}{9}$ &
$-\frac{2}{9}$ & 0 & $\frac{2}{3}$ &
$-\frac{2}{3}$ & 0\tabularnewline \hline $\Xi^{*-}$ &
$-\frac{2}{9}$ & $-\frac{4}{9}$ & 0 &
$-\frac{2}{3}$ & $-\frac{4}{3}$ &
0\tabularnewline \hline $\Omega^{-}$ & 0 &
$-\frac{2}{3}$ & 0 & 0 & -$2$ &
0\tabularnewline \bottomrule[1pt]\bottomrule[1pt]
\end{tabular}
\caption{The coefficients of the loop corrections to the magnetic
moments of the decuplet baryons from Fig. \ref{fig:allloop}(d). The
subscripts ``$\mathcal T$'' and ``$\mathcal N$'' denote the decuplet and octet baryon within
the loop while the superscripts denote the pseudoscalar meson. }
\label{table:beta}
\end{table}

\begin{table}
  \centering
\begin{tabular}{c|c|c|c|c|c|c|c|c|c}
\toprule[1pt]\toprule[1pt] baryons & $\gamma_{a\mathcal T}^{\pi}$ &
$\gamma_{a\mathcal T}^{K}$ & $\gamma_{a\mathcal T}^{\eta}$ &
$\gamma_{a\mathcal N}^{\pi}$ & $\gamma_{a\mathcal N}^{K}$ &
$\gamma_{a\mathcal N}^{\eta}$&$\gamma_{a\mathcal T\mathcal N}^{\pi}$  &
$\gamma_{a\mathcal T\mathcal N}^{K}$ & $\gamma_{a\mathcal T\mathcal N}^{\eta}$\tabularnewline \midrule[1pt] $\Delta^{++}$ &
$\frac{8}{3}$ & $\frac{2}{3}$ &
$\frac{2}{3}$ &
$\frac{2}{3}(b_{D}+3b_{F})$ &
$\frac{2}{3}(b_{D}+3b_{F})$ & 0& $-\frac{4}{3}$ &
$-\frac{4}{3}$ & 0 \tabularnewline \hline
$\Delta^{+}$ & $\frac{13}{9}$ &
$\frac{2}{9}$ & $\frac{1}{3}$ &
$\frac{4}{3}b_{F}$ &
$\frac{2}{3}(b_{D}+b_{F})$ & 0&
$-\frac{4}{9}$ & $-\frac{8}{9}$ & $0$ \tabularnewline \hline
$\Delta^{0}$  & $\frac{2}{9}$ &
$\frac{-2}{9}$ & 0 &
$\frac{-2}{3}(b_{D}-b_{F})$ &
$\frac{2}{3}(b_{D}-b_{F})$ & 0&
$\frac{4}{3}$ &
$-\frac{4}{9}$ & $0$\tabularnewline \hline
$\Delta^{-}$  & $-1$ & $-\frac{2}{3}$
& $-\frac{1}{3}$ &
$\frac{-8}{9}b_{D}$ &
$\frac{4}{9}(b_{D}-3b_{F})$ & 0&
$\frac{4}{3}$ & $0$ & $0$\tabularnewline \hline
$\Sigma^{*+}$  & $\frac{4}{9}$ &
$\frac{14}{9}$ & 0 &
$\frac{1}{9}(-b_{D}+3b_{F})$ &
$\frac{-2}{9}(b_{D}-3b_{F})$ &
$\frac{1}{3}(b_{D}+3b_{F})$&
$0$ &
$-\frac{4}{3}$ & 0 \tabularnewline \hline
$\Sigma^{*0}$ & 0 & 0 & 0 & $-\frac{1}{9}b_{D}$ &
$-\frac{2}{9}b_{D}$ &
$\frac{1}{3}b_{D}$&
$\frac{4}{9}$ &
$-\frac{2}{3}$ & 0 \tabularnewline \hline
$\Sigma^{*-}$ & $-\frac{4}{9}$ &
$\frac{-14}{9}$ & 0 &
$\frac{-1}{9}(b_{D}+3b_{F})$ &
$\frac{-2}{9}(b_{D}+3b_{F})$ &
$\frac{1}{3}(b_{D}-3b_{F})$& $\frac{8}{9}$ &
$\frac{4}{9}$ & 0 \tabularnewline \hline
$\Xi^{*0}$ & $\frac{-2}{9}$ &
$\frac{2}{9}$ & 0 &
$\frac{-2}{3}b_{F}$ &
$\frac{2}{3}b_{F}$ &
$\frac{-2}{3}b_{D}$&
$\frac{2}{9}$ &
$\frac{4}{9}$ &
$\frac{2}{3}$ \tabularnewline \hline $\Xi^{*-}$
& $-\frac{1}{9}$ & $\frac{-14}{9}$ &
$\frac{-1}{3}$ &
$\frac{1}{3}(-b_{D}-b_{F})$ &
$\frac{-2}{3}b_{F}$ &$\frac{1}{3}(b_{D}-3b_{F})$&
$\frac{4}{9}$ &
$\frac{8}{9}$ & 0\tabularnewline \hline
$\Omega^{-}$ & 0 & $\frac{-2}{3}$ &
$-\frac{4}{3}$ & 0 &
$\frac{-4}{9}(b_{D}+3b_{F})$ & 0& 0 &
$-\frac{4}{3}$ & 0\tabularnewline
\bottomrule[1pt]\bottomrule[1pt]
\end{tabular}
\caption{The coefficients of the loop corrections to the magnetic
moments of the decuplet baryons from Fig. \ref{fig:allloop}(a).}
\label{gHQH2}
\end{table}

\begin{table}
  \centering
\begin{tabular}{c|c|c|c|c|c|c|c|c|c|c|c|c}
\toprule[1pt]\toprule[1pt] baryons  & $\gamma_{b}^{\pi}$
& $\gamma_{b}^{K}$ & $\gamma_{b}^{\eta}$ & $\gamma_{e}^{\pi}$
& $\gamma_{e}^{K}$ & $\gamma_{e}^{\eta}$& $\gamma_{f\mathcal T}^{\pi}$ &
$\gamma_{f\mathcal T}^{K}$ & $\gamma_{f\mathcal T}^{\eta}$ &
$\gamma_{f\mathcal N}^{\pi}$ & $\gamma_{f\mathcal N}^{K}$ &
$\gamma_{f\mathcal N}^{\eta}$\tabularnewline \hline
$\Delta^{++}$ & $-b$ & $-b$ & 0 &
$2g_{2}$ & $2g_{2}$ & 0&
$\frac{5}{3}$ & $\frac{2}{3}$ &
$\frac{1}{3}$ & $2$ &
$2$ & 0\tabularnewline \hline
$\Delta^{+}$ &
$-\frac{1}{3}b$ & $-\frac{2}{3}b$ & 0 & $2g_{2}$ &
$\frac{4}{3}g_{2}$ & 0&
$\frac{5}{3}$ & $\frac{2}{3}$ &
$\frac{1}{3}$ & $2$ &
$2$ & 0\tabularnewline \hline
$\Delta^{0}$ & $\frac{1}{3}b$ &
$-\frac{1}{3}b$ & 0 & $2g_{2}$ & $\frac{2}{3}g_{2}$ &
0&
$\frac{5}{3}$ & $\frac{2}{3}$ &
$\frac{1}{3}$ & $2$ &
$2$ & 0\tabularnewline \hline
$\Delta^{-}$ & $b$ & $b$ &
0 & $3g_{2}$ & 0 & 0&
$\frac{5}{3}$ & $\frac{2}{3}$ &
$\frac{1}{3}$ & $2$ &
$2$ & 0\tabularnewline \hline
$\Sigma^{*+}$& $-\frac{2}{3}b$ &
$-\frac{1}{3}b$ & 0 & $\frac{4}{3}g_{2}$ & $2g_{2}$ &
0&
$\frac{8}{9}$ & $\frac{16}{9}$ & 0 &
$\frac{5}{3}$ & $\frac{4}{3}$ &
$1$\tabularnewline \hline
$\Sigma^{*0}$ & 0 & 0 & 0 &
$\frac{4}{3}g_{2}$ & $\frac{4}{3}g_{2}$ & 0&
$\frac{8}{9}$ & $\frac{16}{9}$ & 0 &
$\frac{5}{3}$ & $\frac{4}{3}$ &
$1$\tabularnewline \hline
$\Sigma^{*-}$ & $\frac{2}{3}b$ &
$\frac{1}{3}b$ & 0 & $\frac{4}{3}g_{2}$ & $\frac{2}{3}g_{2}$ &
0&
$\frac{8}{9}$ & $\frac{16}{9}$ & 0 &
$\frac{5}{3}$ & $\frac{4}{3}$ &
$1$\tabularnewline \hline
$\Xi^{*0}$ & $-\frac{1}{3}b$ &
$\frac{1}{3}b$ & 0 & $\frac{2}{3}g_{2}$ & $2g_{2}$ &
0&
$\frac{1}{3}$ & $2$ &
$\frac{1}{3}$ & $1$ &
$2$ & $1$\tabularnewline \hline $\Xi^{*-}$  & $\frac{1}{3}b$ &
$\frac{2}{3}b$ & 0 & $\frac{2}{3}g_{2}$ & $\frac{4}{3}g_{2}$ &
0 & $\frac{1}{3}$ & $2$ &
$\frac{1}{3}$ & $1$ &
$2$ & $1$\tabularnewline \hline $\Omega^{-}$  & 0 & $b$ & 0 & 0 &
$2g_{2}$ & 0& 0 & $\frac{4}{3}$ &
$\frac{4}{3}$ & 0 & $4$ &
0\tabularnewline \bottomrule[1pt]\bottomrule[1pt]
\end{tabular}
\caption{The coefficients of the loop corrections to the magnetic
moments of the decuplet baryons from Figs.  \ref{fig:allloop}(b),
\ref{fig:allloop}(e) and \ref{fig:allloop}(f).} \label{gb}
\end{table}

\end{appendix}

\vfil \thispagestyle{empty}

\end{document}